\documentclass[12pt]{iopart}
\pdfoutput=1
\usepackage{amsfonts}
\usepackage{graphicx}

\usepackage{xcolor}
\usepackage{subfigure}
\usepackage{textcomp}
\usepackage{cite}
\usepackage{float}
\usepackage{soul}
\usepackage{stackrel}

\definecolor{dgreen}{rgb}{0,0.7,0}

\expandafter\let\csname equation*\endcsname\relax
\expandafter\let\csname endequation*\endcsname\relax
\usepackage{amsmath,amssymb}

\pdfoutput=1
\usepackage{esint}

\usepackage{color}
\definecolor{dgreen}{rgb}{0,0.7,0}

\newcommand{\MeijerG}[7]{G \begin{smallmatrix} #1 & #2 \\ #3 & #4 \end{smallmatrix} \left( \begin{smallmatrix} #5 \\ #6 \end{smallmatrix} \middle\vert #7 \right) }
  
\begin{document}

\title[]{Random acceleration process under stochastic resetting}

\author{Prashant Singh}

\address{International Centre for Theoretical Sciences, Tata Institute of Fundamental
Research, Bengaluru 560089, India}
\ead{prashant.singh@icts.res.in}
\vspace{10pt}
%\begin{indented}
%\item[]August 2017
%\end{indented}

\begin{abstract}
We consider the motion of a randomly accelerated particle in one dimension under stochastic resetting mechanism. Denoting the position and velocity by $x$ and $v$ respectively, we consider two different resetting protocols - (i) complete resetting: here both $x$ and $v$ reset to their initial values $x_0$ and $v_0$ at a constant rate $r$, (ii) partial resetting: here only $x$ resets to $x_0$ while $v$ evolves without interruption. For complete resetting, we find that the particle attains stationary state in both $x$ and $v$. We compute the non-equilibrium joint stationary state of $x$ and $v$ and also study the late time relaxation of the distribution function. On the other hand, for partial resetting, the joint distribution is always in the transient state. At large $t$, the position distribution possesses a scaling behaviour $(x/ \sqrt{t})$ which we rigorously derive. Next, we study the first passage time properties with an absorbing wall at the origin. For complete resetting, we find that the mean first passage time is rendered finite by the resetting mechanism. We explicitly derive the expressions for the mean first passage time and the survival probability at large $t$. However, in stark contrast, for partial resetting, we find that resetting does not render finite mean first passage time. This is because even though $x$ is brought to $x_0$, the large fluctuation in $v$ ($\sim \sqrt{t}$) can take the particle substantially far from the origin. All our analytic results are corroborated by the numerical simulations.
 
\end{abstract}

\section{Introduction}
\label{intro}
Stochastic resetting refers to an intermittent interruption of a dynamical process after which the process starts anew. This simple act of stopping a process and restarting leads to a plethora of interesting phenomena like - existence of non-trivial stationary states, temporal relaxations and dynamical phase transitions, finite mean first passage times etc \cite{Bartumeus2009, Luby1993,Montanari2002, Reuveni2014, Rotbart2015, Roldan2016,Kusmierz2014, Bressloff2020, Cortes2017, Chechkin2018, Pal2019,Bela2018, Evans2011, Pal2015, Majumdar2015,Ray2020, EvansSatya2011, EvansSatya2013, PalShlomi2017, PalPrasad2019, Bruyne2020, PrasadPal2019}. Resetting has been very useful as an efficient search strategy in a wide range interdisciplinary applications like foraging \cite{Bartumeus2009}, computer science \cite{Luby1993,Montanari2002}, biological and chemical processes \cite{Reuveni2014, Rotbart2015, Roldan2016}, search process \cite{Kusmierz2014, Bressloff2020, Cortes2017, Chechkin2018, Pal2019,Bela2018} and so on. Not only in the theoretical research front, resetting has been realised recently in optical tweezers based experiments \cite{Friedman2020, Besga2020}. This experimental progress, in addition to the widespread applications, have led to a surged interest and exploration of this field in the recent years (see \cite{Evans2020} for a review).

\begin{figure}[]
\includegraphics[scale=0.3]{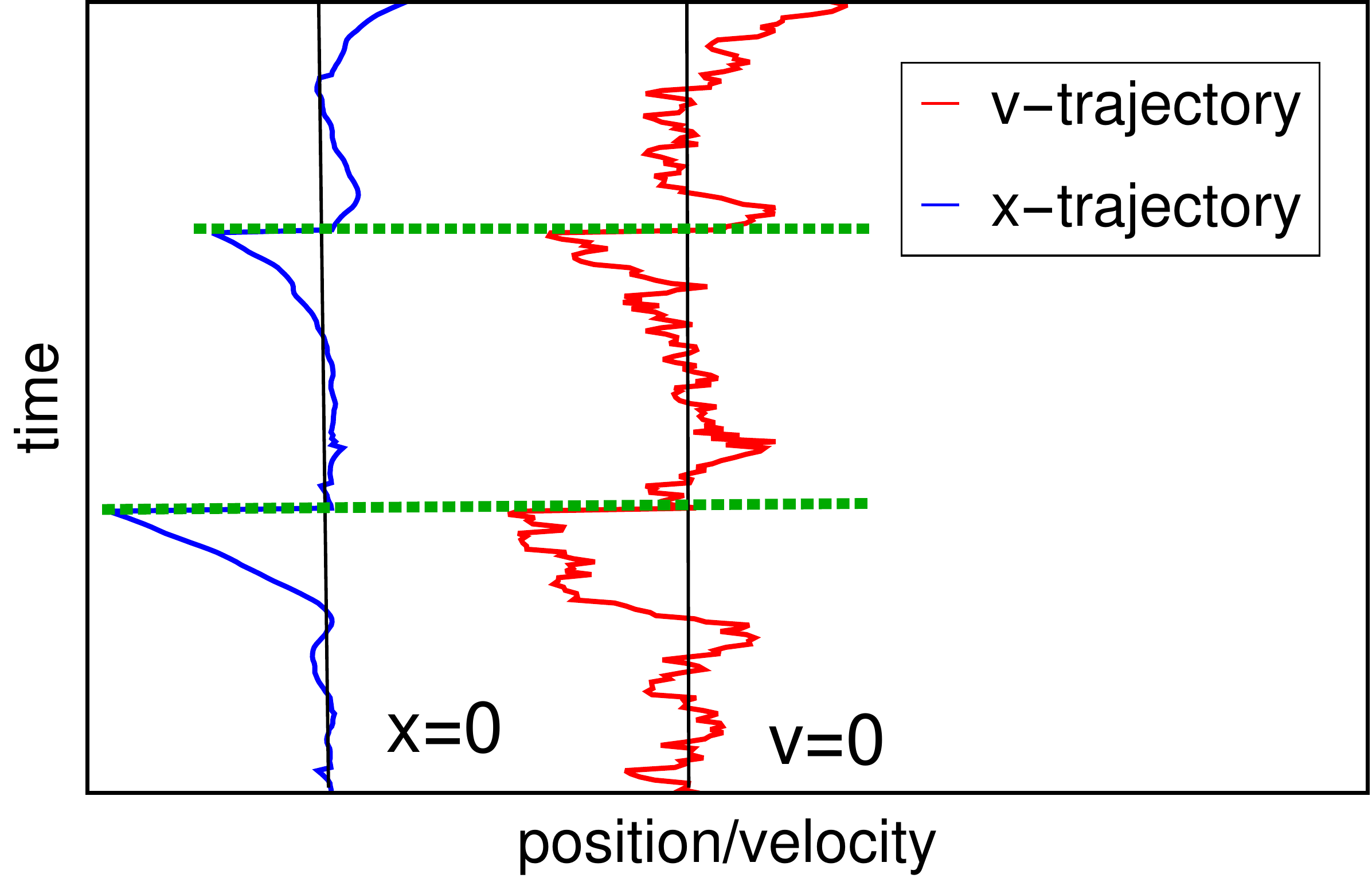}
\includegraphics[scale=0.27]{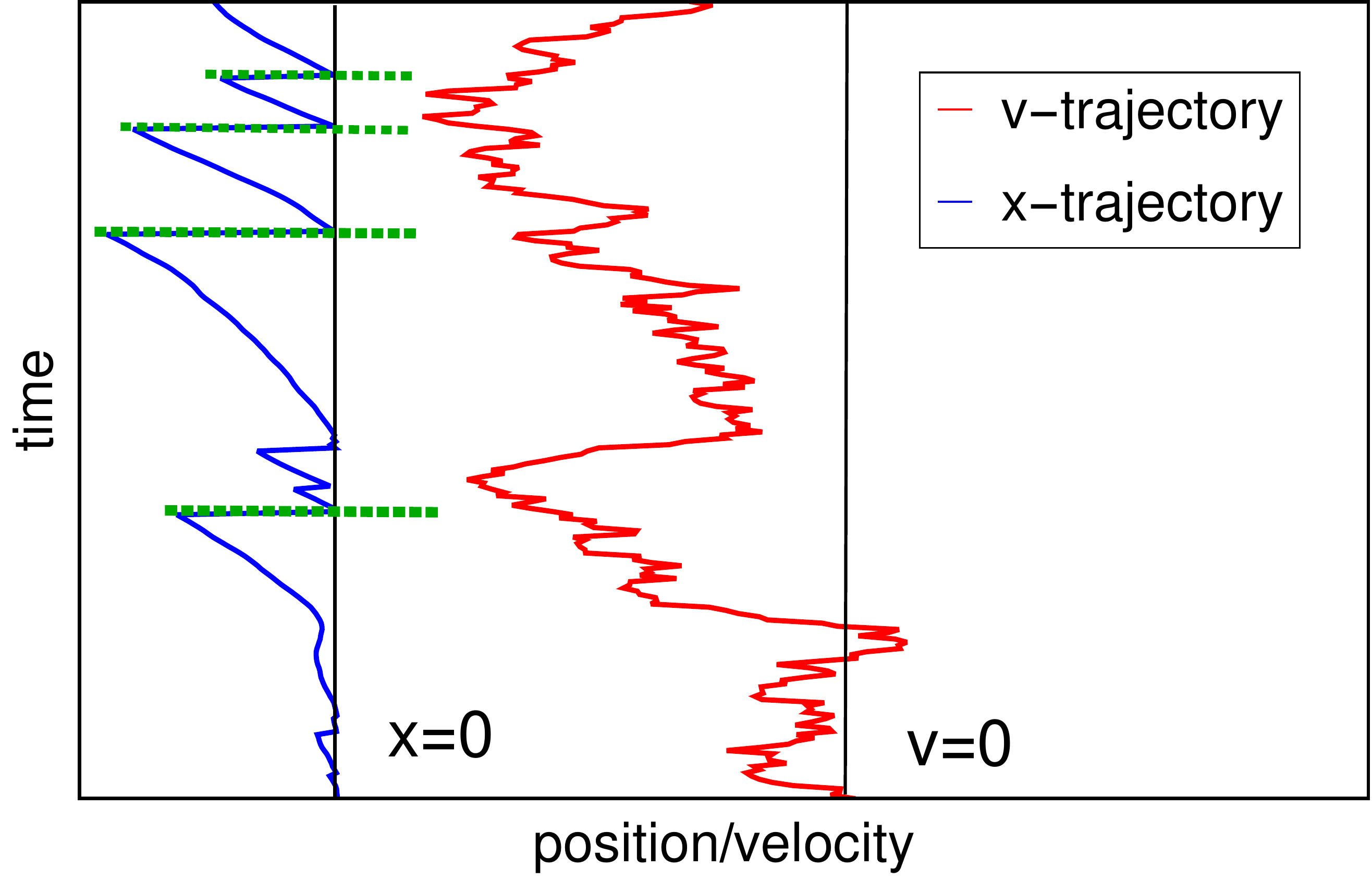}
\centering
\caption{In left panel, we have trajectories of $x$ and $v$ when both resets (shown by dashed green line) to their respective initial values $x_0 = 0$ and $v_0 = 0$ at a constant rate $r$. The dynamics, in-between successive resetting events, is governed by Eqs. \eqref{langevin-eq-2}. In right panel, the same is shown for the case of partial resetting where $x$ is reset while $v$ evolves without interruption.}    
\label{schematic-traj}
\end{figure}

The paradigm of the subject is perhaps the effect of resetting on a single diffusing particle and myriad of its properties are known. Some examples are - stationary probability distributions and relaxations \cite{Evans2011, Pal2015, Majumdar2015,Ray2020}, optimised first passage properties \cite{EvansSatya2011, EvansSatya2013, PalShlomi2017, PalPrasad2019, Bruyne2020, PrasadPal2019}, path functionals \cite{Meylahn2015, Hollander2019, PalChaterjee2019}. It has also been studied in the context of other stochastic processes like RTP \cite{EvansMajumdar2018}, telegraphic process \cite{Masoliver2019}, anomalous diffusion \cite{MasoliverMontero2019, Mend2016,Kusmi2019}, fractional diffusion \cite{santos2019}, L\'{e}vy flights \cite{Kusmierz2014, KusmierzNowak2015}, scaled Brownian motion \cite{Bodrova2019}, underdamped Brownian motion \cite{Gupta2019}, quantum systems \cite{Mukherjee2018, Rose2018} and extended systems with many degrees of freedom \cite{SGuptaMajumdar2014,Basu2019, Karthika2020,Sadekar2020}. Stochastic thermodynamics of resetting has also been studied in \cite{PalRaghav2017, Gupta2020}. Although most of these studies have focussed on exponential resetting to the initial configuration, there have also been some studies with non-exponential (like power-law) resetting \cite{Nagar2016,GuptaNagar2016,Eule2016,Bodr0Sok2020}, non-Markovian resetting \cite{MajumdarSabhapandit2015, Boyer2019,PalKundu2016,Kusm2019}.\\
In this paper, we consider the stochastic process in which a particle is randomly accelerated in one dimension. The Langevin equation of the particle is given by,
\begin{align}
\frac{d^2 x}{d t^2} = \eta (t),
\label{langevin-eq-1}
\end{align}
where $\eta (t)$ is the Gaussian white noise with $\langle \eta (t) \rangle=0$ and $\langle \eta(t) \eta(t') \rangle = 2 D \delta (t-t')$. For simplicity we choose $D=1$. This is a non-Markov process as indicated by the second order time derivative in Eq. \eqref{langevin-eq-1}. However if one defines velocity $v = \frac{d x}{dt}$, then the process becomes Markovian in the phase space $(x,v)$. The Langevin Eq. \eqref{langevin-eq-1} can now be rewritten as,
\begin{align}
\frac{d v}{dt} = \eta(t),~~~~~\frac{d x}{dt} = v.
\label{langevin-eq-2}
\end{align}
Here we study the random accerleration process (henceforth RAP) under the effect of resetting characterised by a constant rate $r$. We use the renewal equation scheme to study the probability distribution as well as the persistent properties of the particle \cite{EvansMajumdar2018}. For RAP, resetting happens in $(x,v)$ phase space. This gives rise to various possibilities in which $x$ as well as $v$ can reset. In this paper we consider two different resetting protocols - (i) both $x$ and $v$ are reset to their initial values $x_0$ and $v_0$ at a constant rate $r$, (ii) only $x$ is reset to its initial value $x_0$, while $v$ evolves without interruption. For convenience we will refer to (i) as complete resetting and (ii) as partial resetting. The dynamics, in-between successive resetting events, is governed by the Langevin Eqs. \eqref{langevin-eq-2}. In Figure \ref{schematic-traj}, we have shown the trajectories of $x$ and $v$ corresponding to the two resetting protocols. The update rules for $x$ and $v$ in the small time interval $t$ to $t+dt$ are given by,

\begin{align}
 \text{Protocol-} (\text{i})
    \begin{cases}
    \begin{split}
      v(t+dt)&= v(t) + \eta (t) dt ~~~~\text{ with probability }1-r dt, \\
      &= v_0 ~~~~~~~~~~~~~~~~~~ \text{ with probability }r dt.\\
      x(t+dt)&= x(t) + v(t) dt ~~~~\text{ with probability }1-r dt, \\
      &= x_0 ~~~~~~~~~~~~~~~~~~ \text{ with probability }r dt.\\
      \end{split}
    \end{cases}
\label{protocol-eq-1} 
\end{align} 

\begin{align}
 \text{Protocol-} \text{(ii)}
    \begin{cases}
    \begin{split}
      v(t+dt)&= v(t) + \eta (t) dt ~~~~\text{ with probability }1, \\
      x(t+dt)&= x(t) + v(t) dt ~~~~\text{ with probability }1-r dt, \\
      &= x_0 ~~~~~~~~~~~~~~~~~~ \text{ with probability }r dt.\\
      \end{split}
    \end{cases}
\label{protocol-eq-2} 
\end{align} 
Our study reveals that while the particle exhibits stationary state as $t \to \infty$ for the complete resetting protocol, the probability distribution is always in the transient state for the partial case. In fact, the latter protocol does not render finite mean first passage time in contrast to the complete resetting where we find that the mean first passage time is finite. At this point, we emphasise that resetting with two dynamical variables was also studied for run and tumble particle \cite{EvansMajumdar2018} and underdamped Brownian motion \cite{Gupta2019}. The velocity variable $v$ for run and tumble particle can take two possible values. For underdamped Brownian motion, $v$ exhibits a stationary state in absence of resetting. On the other hand, velocity $v$ in RAP is unbounded in absence of resetting. The fluctuations in $v$ grow with time as $\sim \sqrt{t}$. This, as we show in this paper, gives rise to effects that are strikingly different than in \cite{EvansMajumdar2018, Gupta2019}.

The paper is organised as follows. We begin with a brief review of RAP in absence of resetting in sec.~\ref{prob-recap}. We study the distribution for complete resetting in sec.~\ref{prob-x-v} and for partial resetting in sec.~\ref{prob-x}. After studying probability distribution, we next investigate the first passage time properties. We briefly summarise the first passage properties of RAP in absence of resetting in sec.~\ref{surv-no-reset}. We perform computations of survival probability with an absorbing barrier at $x=0$ for complete resetting in sec.~\ref{survv-x-v} and for partial resetting in sec.~\ref{surv-x}. Finally, we conclude in sec.~\ref{conclusion}.

\section{Random acceleration process in absence of resetting}   
\label{prob-recap}
We begin by reviewing some known results for RAP (in absence of resetting) which will prove relevant for the current work. At first, we summarize some known results for RAP which has found applications in many areas of physics (see \cite{Burkhardt2016} for review). For example semi-flexible polymer in a narrow cylindrical channel is described by RAP \cite{Burkhawdt1993, Burkhawdt1997}. Also the spatial persistence of fluctuating interface with $z = 4$ is described by RAP \cite{MajumdarBray2001}. Other applications include crack propagation in elastic media \cite{Schwarz2001}, statistical properties of Burgers equation with Brownian initial velocity \cite{Valageas2009}. Given these widespread applications, alongwith being the poster boy for non-Markov processes, RAP has been extensively studied and a variety of its properties are known. The list includes - first passage properties and persistence exponents \cite{Burkhardt2016, Burkhawdt1993, Mckean1963, Goldman1971, Marshall1985,ScherMajumdar2013}, convex hull \cite{Reymbaut2011}, extreme value statistics \cite{Burkhardt2008}, distribution for time to reach maximum distance \cite{MajumdarRosso2010}, last passage time distribution \cite{Lachal1994}, occupation time statistics \cite{JoelBout2016, Burkhardt2017}, exit time statistics \cite{Kotsev2005, Burkhardt2007}, tagged particle statistics in single file process \cite{WBurkhardt2019}, functional distributions \cite{Sinai1992}, partial survival probability \cite{WTBurkhardt2000,SmedtDe2001} etc.

As discussed before, the process is Markovian in the $(x,v)$ phase space. Denoting by $P_0(x,v,t|x_0,v_0)$ the joint probability distribution in an infinite line $-\infty < x < \infty$ for the particle to be at $(x,v)$ in time $t$ given that it was initially at $(x_0, v_0)$, one can write forward master equation \cite{Burkhawdt1993} as, 
\begin{align}
\frac{\partial P_0}{\partial t}=\frac{\partial^2 P_0}{\partial v^2}-v \frac{\partial P_0}{\partial x}.
\label{fokker}
\end{align} 
The subscript $0$ in $P_0(x,v,t|x_0,v_0)$ indicates the process in absence of resetting. The initial condition is
\begin{align}
P_0(x,v,0|x_0,v_0) = \delta (x-x_0) \delta (v-v_0),
\label{initial-dist}
\end{align}
and the boundary conditions are $P_0(x,v,t|x_0,v_0)$ should remain finite as $x \to \pm \infty$ and/or $v \to \pm \infty$. One can, in principle, solve the master Eq. \eqref{fokker} alongwith the boundary conditions to get $P_0(x,v,t|x_0,v_0)$. However, we follow a well known property of Gaussian variables to write the distribution. Note that $x$ and $v$ as indicated by the Langevin Eqs. \eqref{langevin-eq-2} are Gaussian variables. We can write the joint distribution of $x$ and $v$ as,
\begin{align}
P_0(x,v,t|x_0, v_0) = \mathcal{N} ~\text{exp}\left(-\frac{1}{2}X^T \mathcal{C}^{-1} X \right),
\label{formula-eq-1}
\end{align}
where $\mathcal{N}$ is the constant fixed by the normalisation condition $\int_{-\infty}^{\infty} dx~ dv ~P_0(x,v,t|x_0,v_0)=1 $ and $X^T = \left(x-\langle x \rangle~~~ v-\langle v \rangle \right)$ and $\mathcal{C}$ is the covariance matrix defined as,
\begin{align}
\mathcal{C} = 
\begin{pmatrix}
&\langle x^2 \rangle -\langle x \rangle^2 & \langle x v \rangle -\langle x \rangle \langle v \rangle\\
&\langle x v \rangle -\langle x \rangle \langle v \rangle &\langle v^2 \rangle -\langle v \rangle^2
\end{pmatrix}
.
\label{covar-eq-1}
\end{align}
It is straight forward to show from the Langevin Eqs. \eqref{langevin-eq-2} that
\begin{align}
&\langle x \rangle = x_0+v_0 t , \nonumber \\
& \langle v \rangle = v_0, \nonumber \\
&\mathcal{C} = 
\begin{pmatrix}
&2t^3/3 & t^2\\
&t^2 &  2t
\end{pmatrix}
.
\label{covar-eq-2}
\end{align}
Inserting these expressions in Eq. \eqref{formula-eq-1} one gets the expression of $P_0(x,v,t|x_0,v_0)$ which reads,
\begin{align}
P_0(x,v,t|x_0,v_0)=\frac{\sqrt{3}}{2 \pi t^2}\text{exp}\left[-\frac{3}{t^3} \big\{(x-x_0-v t)(x-x_0-v_0 t)+\frac{t^2}{3}(v-v_0)^2 \big\} \right].
\label{fokk_sol-50}
\end{align}
To get the distribution of $x$, we integrate $P_0(x,v,t|x_0,v_0)$ over all $v$ as,
\begin{align}
P_0(x,t|x_0,v_0)= \sqrt{\frac{3}{4 \pi t^3}} \text{exp} \left[ -\frac{3}{4 t^3} (x-v_0 t-x_0)^2\right].
\label{fokk_sol-eq2}
\end{align}
For latter calculations, we also need the Laplace transform of $P_0(x,t|x_0,v_0)$ in $t$. The Laplace transform of the distribution with respect to $t$ is defined as,
\begin{align}
\bar{P}_0(x,v,s|x_0,v_0) = \int_{0}^{\infty} dt ~e^{-s t} P_0(x,v,t|x_0,v_0).
\label{lap-def}
\end{align} 
Following \cite{Burkhawdt1993}, one can write $\bar{P}_0(x,v,s|x_0,v_0)$ as,
\begin{align}
\bar{P_0}(x,v,s|x_0,v_0)&= \int_{0}^{\infty} dF \frac{e^{-F \mid x-x_0 \mid}}{F^{1/3}} \text{Ai}\left[-\text{sgn}(x-x_0) v F^{1/3}+\frac{s}{F^{2/3}} \right]  \nonumber \\
&~~~~~~~~~~~~~~~~~~~~~\text{Ai}\left[-\text{sgn}(x-x_0) v_0 F^{1/3}+\frac{s}{F^{2/3}} \right],
\label{fokk_sol1}
\end{align}
where the signum function $\text{sgn}(x)$ is $1$ for $x>0$, $-1$ for $x<0$ and $0$ for $x=0$. Also here, Ai$(z)$ is the Airy function. Integrating the joint distribution over all $v$ and using $\int_{-\infty}^{\infty} dz \text{Ai}(z)=1$ , we get the Laplace transform of the distribution for $x$ as,
\begin{align}
\bar{P_0}(x,s|0,v_0)= \int_{0}^{\infty} dF \frac{e^{-F \mid x \mid}}{F^{2/3}} \text{Ai}\left[-\text{sgn}(x)v_0F^{1/3}+\frac{s}{F^{2/3}} \right],
\label{fokk_sol}
\end{align}
where we have taken $x_0=0$ for simplicity. In what follows, we use the joint distribution in Eq. \eqref{fokk_sol-50} and the Laplace transforms in Eqs. \eqref{fokk_sol1} and \eqref{fokk_sol} to analyse the process in presence of the resetting.
 \begin{figure}[t]
\includegraphics[scale=0.3]{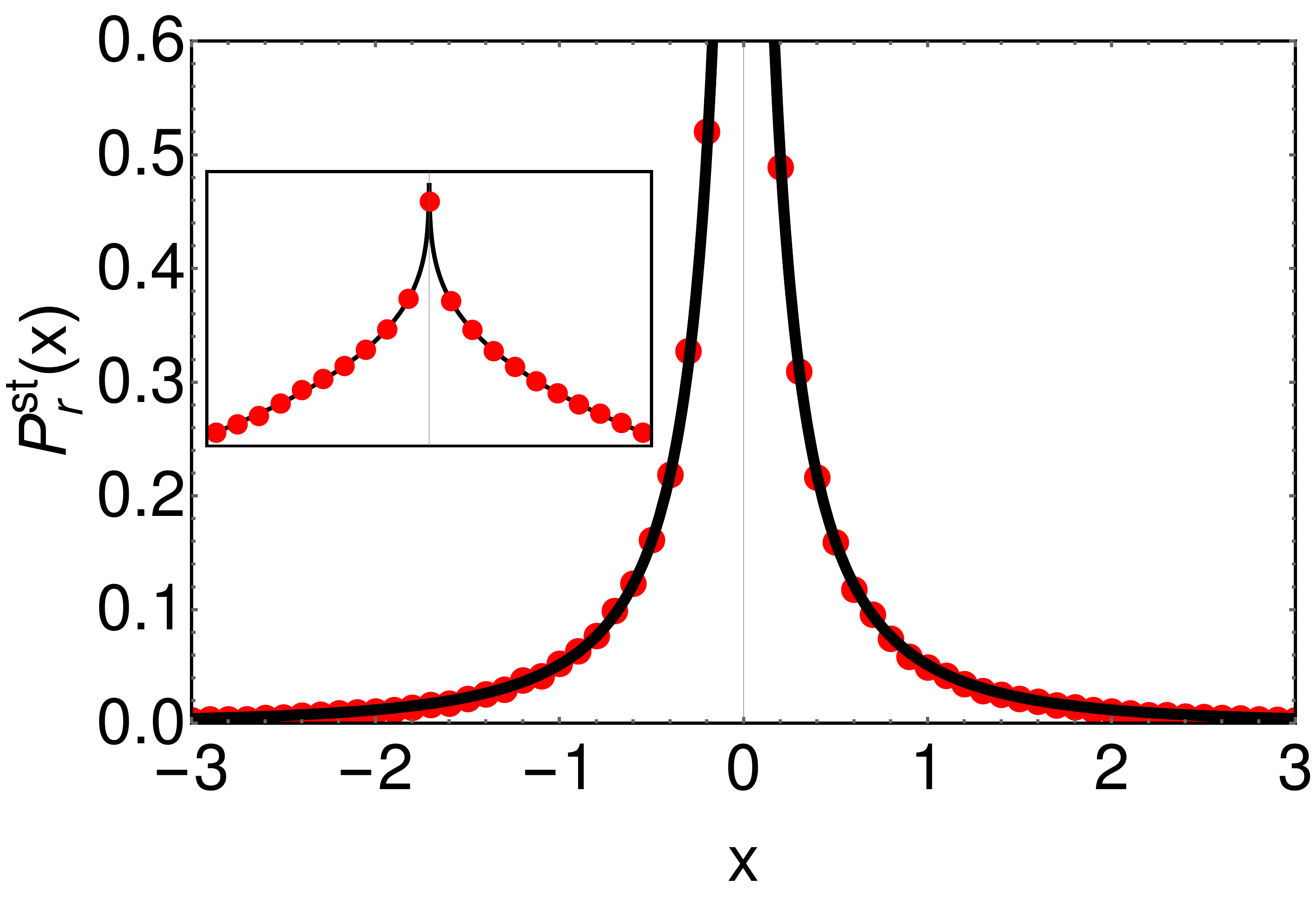}
\centering
\caption{Stationary distribution $P^{st}_r(x)$ in Eq.\eqref{reset_et_3} for complete resetting is plotted (shown by black solid line) and compared with the simulation of Langevin equation (shown by red filled circles) in Eq. \eqref{langevin-eq-2}. We have chosen $r=2$. The simulation result is obtained for $t=10$. \textit{Inset:} Shows the same plot in log scale.}    
\label{steadypic1}
\end{figure}

\section{Probability distribution with resetting}
\label{prob-resett}
\subsection{When both $x$ and $v$ are reset (complete resetting)}
\label{prob-x-v}
We start with the complete resetting protocol where $x$ and $v$ evolve according to the update rules in Eqs. \eqref{protocol-eq-1}. Both $x$ and $v$ reset to their initial values $x_0$ and $v_0$ respectively at a constant rate $r$. For simplicity, we choose $(x_0,v_0)=(0,0)$. Denoting the probability distribution by $P_r(x,v,t|0,0)$, we use last renewal equation scheme to compute the probability distribution. Renewal equation approach has been also used for other stochastic processes under resetting \cite{EvansMajumdar2018, Gupta2019}. Under this approach the distribution $P_r(x,v,t|0,0)$ is given by,
\begin{align}
P_r(x,v,t|0,0)&=e^{-r t} P_0(x,v,t|0,0)+r \int_{0}^{t} d \tau e^{-r \tau}  P_0(x,v, \tau|0,0) \int dx' dv'P_r(x', v', t-\tau|0,0), \nonumber\\
&=e^{-r t} P_0(x,v,t|0,0)+r \int_{0}^{t} d \tau e^{-r \tau}  P_0(x,v, \tau|0,0) .
\label{reset_et_1}
\end{align}
The equation in the first line has two contributions. The first contribution comes from those paths for which there has not been any resetting event till time $t$. This happens with a probability $e^{-r t}$. The second contribution arises from all those paths for which last reset happens at time $t-\tau$ when the particle is at $(x',v')$. In the remaining interval $t-\tau$ to $t$, the particle does not experience any resetting event as indicated by $e^{-r \tau}$.  In going from first line to the second line, we have used the normalisation condition which gives $\int ~dx ~dv ~P_{r}(x,v,t|0,0)=1$.\\
\textit{Stationary state:} The stationary distribution is obtained by taking $t \to \infty$ limit of $P_r(x,v,t|0,0)$ in Eq. \eqref{reset_et_1}. The first term in Eq. \eqref{reset_et_1} gives zero contribution in this limit while the second term gives non-vanishing contribution to the stationary state. The second term in the $t \to \infty$ limit is just the Laplace transform of $P_0(x,v,t|0,0)$ with respect to $t$ (see Eq. \eqref{lap-def}). Therefore the joint stationary distribution of $(x,v)$ reads, 

\begin{align}
P_r^{st}(x,v)=r \bar{P}_0(x,v,r|0,0),
\label{reset_et_2}
\end{align}
where $\bar{P}_0(x,v,r|0,0)$ is given by Eq. \eqref{fokk_sol1}. Integrating the joint distribution over all $v$  and using $\int_{-\infty}^{\infty} dz \text{Ai}(z)=1$ , we get the stationary distribution of $x$,
\begin{align}
P_r^{st}(x)&=r\int_{-\infty}^{\infty} dv \bar{P}_0(x,v, r|0,0), \label{reset-et-3p} \\
&=r\int_{0}^{\infty} dF \frac{e^{-F \mid x \mid}}{F^{2/3}} \text{Ai}\left[\frac{r}{F^{2/3}} \right],  \label{reset-et-300p}\\
&=r^{3/2}\left[-\sqrt{3}~ _0 F_3 \left(;\frac{1}{2},\frac{5}{6},\frac{7}{6};\frac{r^3 x^2}{36} \right)-\frac{|x|r^{3/2}}{4}~ _0 F_3 \left(;\frac{4}{3},\frac{3}{2},\frac{5}{3};\frac{r^3 x^2}{36} \right) \right. \nonumber \\
&~~~~~~~~~~~~~ \left.\frac{2 \pi}{3 \sqrt{3}} \left\{\mathcal{X}_{1/3}\left(\frac{2 r^{3/4} \sqrt{|x|}}{\sqrt{3}} \right)+\mathcal{Y}_{1/3}\left(\frac{2 r^{3/4} \sqrt{|x|}}{\sqrt{3}} \right) \right\} \right],
\label{reset_et_3}
\end{align}
where $_0F_3$ stands for the generalised hypergeometric function and $\mathcal{X}_{\nu}(z)$ and $\mathcal{Y}_{\nu}(z)$ are defined in terms of Kelvin's bei$_{\nu}$ and ber$_{\nu}$ functions \cite{Kelvin} as,
\begin{align}
&\mathcal{X}_{\nu}(z)=\text{bei}_{\nu}(z)^2+\text{bei}_{-\nu}(z)^2, \nonumber \\
&\mathcal{Y}_{\nu}(z)=\text{ber}_{\nu}(z)^2+\text{ber}_{-\nu}(z)^2.
\end{align}
In going from Eq. \eqref{reset-et-300p} to \eqref{reset_et_3}, we have performed the integration using \textit{Mathematica}. In Figure \ref{steadypic1}, we have plotted the stationary distribution in Eq. \eqref{reset_et_3} and compared it with the numerical simulation. We find an excellent agreement between them. From the expression of $P_r^{st}(x)$, we observe that the length scale over which distribution decays spatially is $l_{p} = r^{-3/2}$. To gain more insights into the distribution, we now look at various limits of $P_r^{st}(x)$ in $x/l_p$. For large $|x|$, it is more convenient to study asymptotics via Eq. \eqref{reset-et-300p} which indicates that the major contribution to the integral will come from small $F$. As the argument of Airy function for small $F$ becomes very large, we use the asympotic form of Ai$(z)$ for large $z$. One then obtains the large $|x|$ behaviour of $P^{st}_r(x)$. The details of this calculation is relegated to \ref{assy-x-pr}. On the other hand for small $|x|/l_p$, we take the leading order term of Eq. \eqref{reset_et_3}. The final approximate expressions for $P_r^{st}(x)$ reads, 

\begin{align}
P_r^{st}(x) &\simeq \frac{\Gamma(1/6) \Gamma(5/6) }{4 \pi} \frac{r^{3/4}}{\sqrt{\mid x \mid}} e^{-\sqrt{\frac{8}{3} r^{3/2} \mid x \mid}},~~~~~~~~~r^{3/2}|x| \to \infty, \label{reset_et_4} \\
& \simeq \frac{0.9511 r}{|x |^{1/3}}, ~~~~~~~~~~~~~~~~~~~~~~~~~~~~~~~~~~~~~~~~r^{3/2}|x| \to 0.
\label{reset_et_478}
\end{align}

It is worth remarking for a diffusing particle that the stationary distribution possesses finite value at the resetting position (which in our case is origin). On the other hand, the stationary distribution for RAP diverges at the resetting position with power-law divergence $\left(|x|^{-\frac{1}{3}} \right)$ as $|x| \to 0$. The divergence can also be seen in Figure \ref{steadypic1} (see inset) which has a cusp as $|x| \to 0$.

\begin{figure}[t]
\includegraphics[scale=0.25]{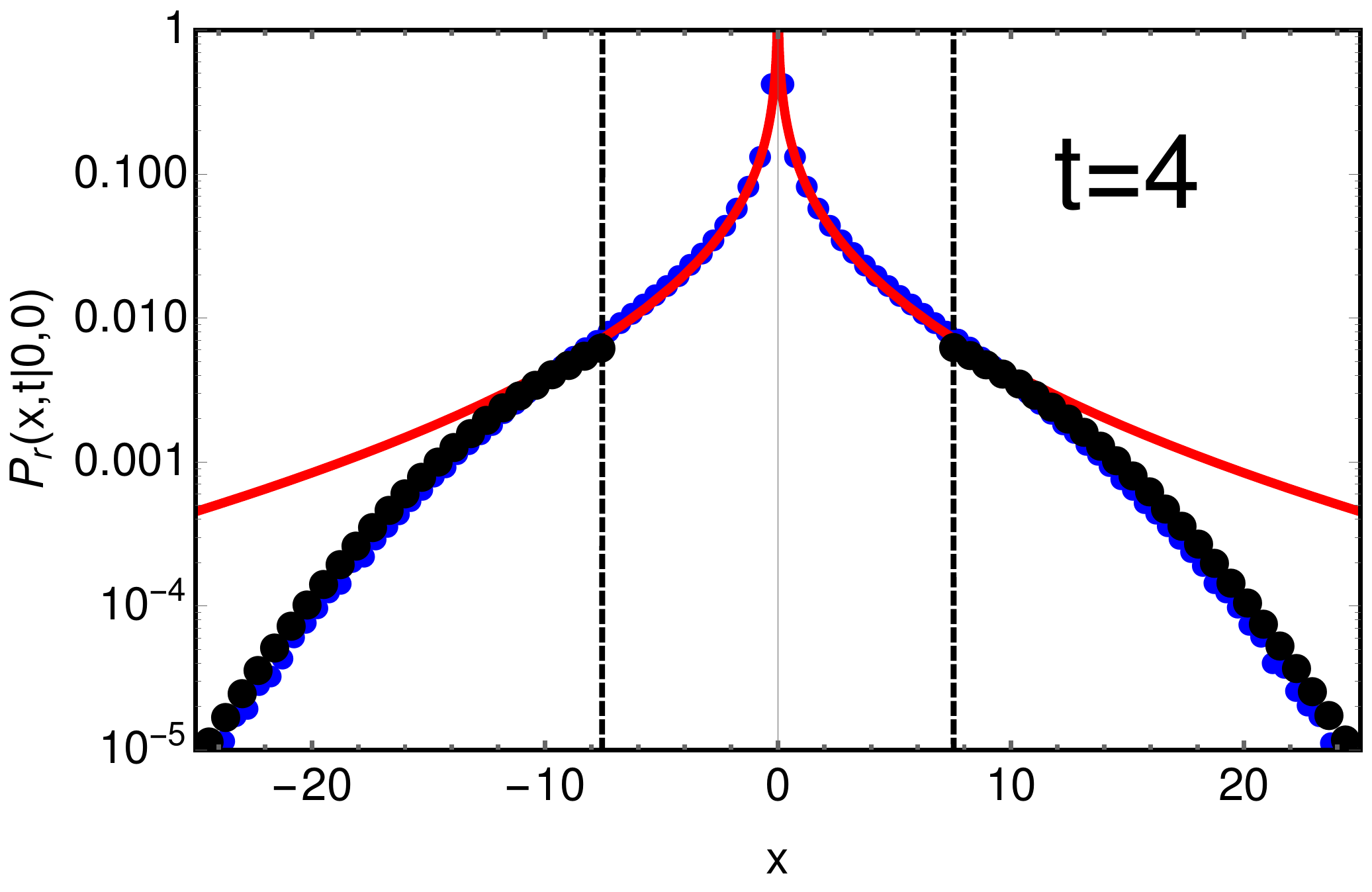}
\includegraphics[scale=0.25]{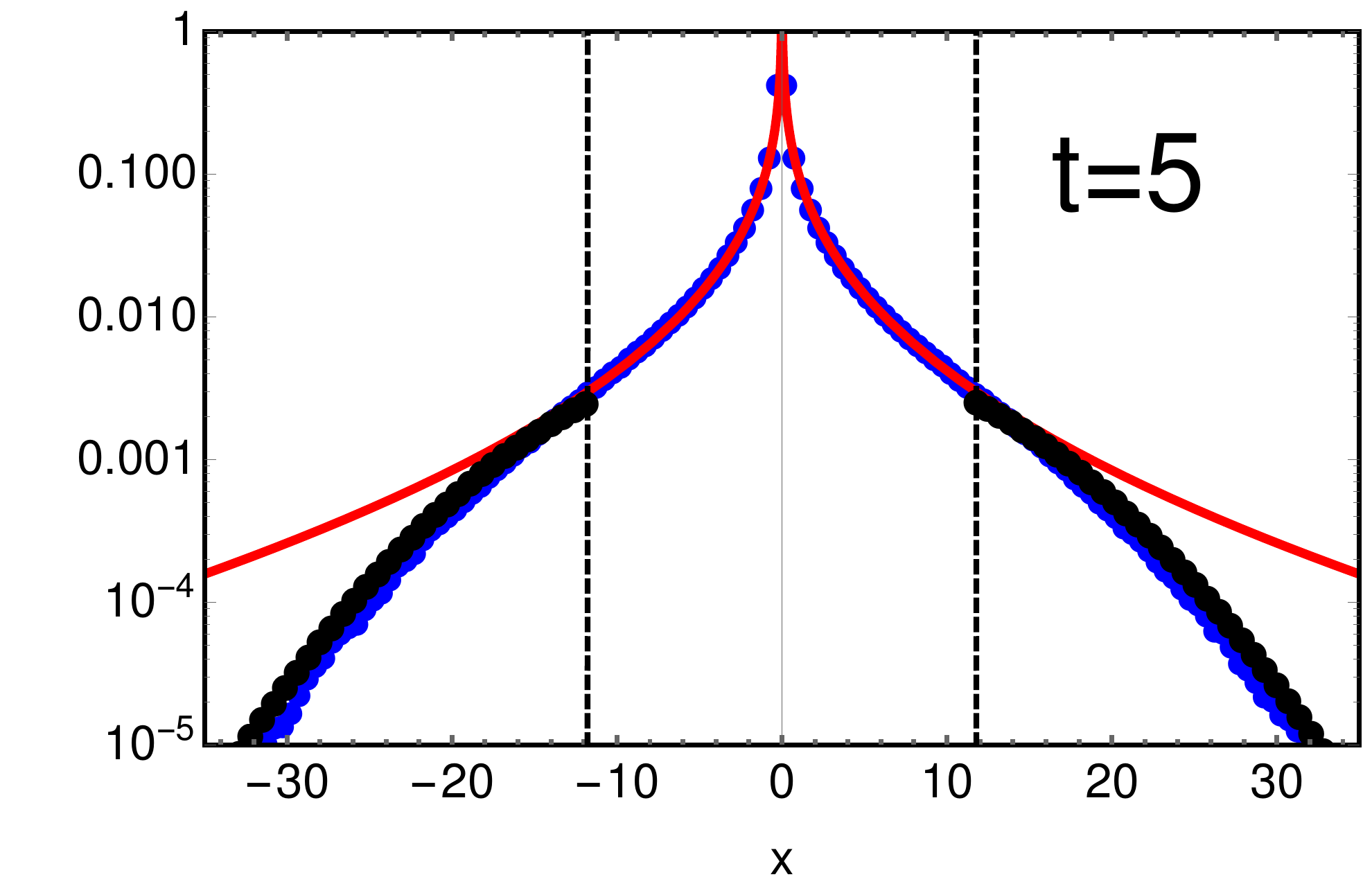}
\includegraphics[scale=0.25]{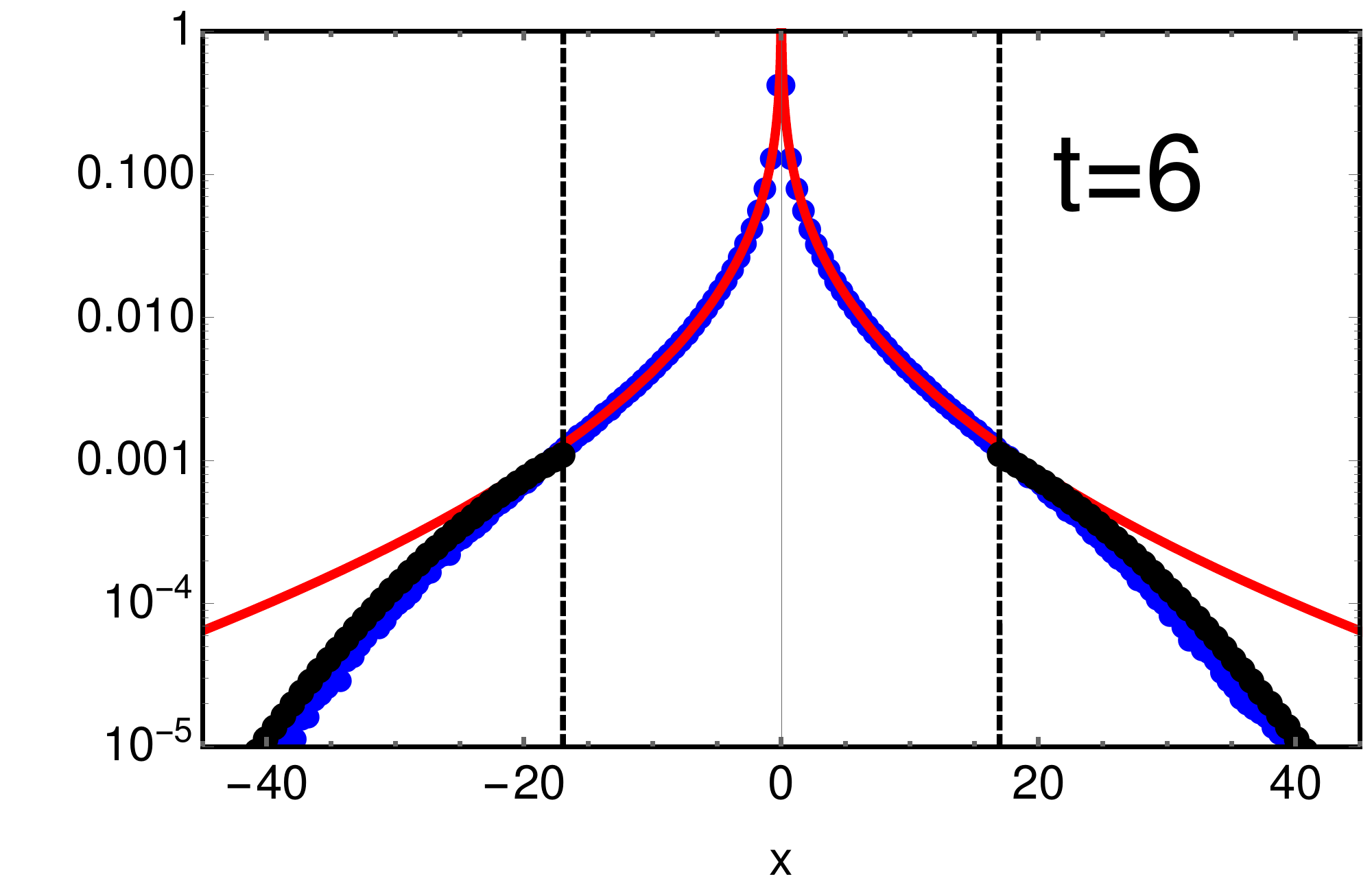}
\centering
\caption{Temporal relaxation of the probability distribution when both $(x,v)$ is reset to $(0,0)$ at a constant rate $r$. The simulation results (shown by blue filled circles) are compared with the analytical results for staionary state (shown by red solid line) in Eq. \eqref{reset_et_3} and the large deviation form (shown by black filled circles) in Eq. \eqref{relax_et_4} for three different times. We have numerically fixed the co efficient in Eq. \eqref{relax_et_4}. For $|x|<x^* $, stationary state has already been achieved while for $|x|> x^* $ the system is still in transient state and yet to equilibrate. The front separating two regimes are shown by the shown by dashed vertical line. We obtain $x^* = 7.54,~11.78,~16.9706$ respectively for $t=4,~5,~6$ from simulations which is consistent with $x^* = \frac{2}{3}\sqrt{r} t^2$ obtained analytically. We have taken $r=0.5$. }    
\label{relax-et-1}
\end{figure} 

\textit{Late time relaxation:} After studying the stationary state, we next turn to the relaxation properties of the probability distribution. For Brownian particle, the relaxation properties were studied in \cite{Majumdar2015,PalKundu2016}. It was shown that while the distribution has attained steady state for $|x| < \sqrt{4r} t$, it was still in the transient state for $|x| >\sqrt{4r} t$. Here we use the same method \cite{Majumdar2015,PalKundu2016} to study the relaxation properties of RAP under the effect of complete resetting. To begin with, we rewrite the last renewal Eq. \eqref{reset_et_1} (after integrating over $v$) as, 
\begin{align}
P_r(x,t|0,0)=e^{-r t} P_0(x,t|0,0)+r \int_{0}^{t} d \tau e^{-r \tau}  P_0(x, \tau|0,0) .
\label{relax_et_1}
\end{align}
Assuming at large $t$, the second term dominates,
\begin{align}
P_r(x,t|0,0)\simeq r \int_{0}^{t} d \tau e^{-r \tau}  P_0(x, \tau|0,0).
\label{relax_et_2}
\end{align}
Inserting $P_0(x,\tau|0,0)$ from Eq. \eqref{fokk_sol-eq2} and changing the variable $\tau = t w$ in Eq. \eqref{relax_et_2}, one gets,
\begin{align}
P_r(x,t|0,0)\simeq \frac{r}{\sqrt{t}} \int_{0}^{1} d w \sqrt{\frac{3}{4 \pi w^3}} e^{-t \Phi\left(w, \frac{x}{t^2} \right)} ,
\label{relax_et_3}
\end{align} 
where $\Phi(w,y)=r w+\frac{3 y^2}{4 w^3}$. For large $t$, we use the saddle point approximation method to evaluate the integral in Eq. \eqref{relax_et_3}. The integral at large $t$ will be dominated by the minimum of $\Phi(w,y)$ in $w$ for fixed $y$. It is easy to check that mimimum of $\Phi(w,y)$ occurs at $w=w^*=\left(\frac{9 y^2}{4 r} \right)^{1/4}$ provided $w^*<1$. On the other hand if $w^* >1$, then $w^*$ lies outside the domain of integration in Eq. \eqref{relax_et_3}. In this case we observe minimum of $\Phi(w,y)$ will occur at $w=1$. Physically this contribution arises from those trajectories for which the particle has not reset. 
\begin{align}
\text{min}\left[ \Phi(w,y)\right] & = \Phi(w^*,y), ~~~~~~\text{~if }w^*<1, \nonumber \\
& = \Phi(1,y),~~~~~~~~ \text{~if }w^*\geq 1.
\label{relax_et__871}
\end{align}
The result is that $P_r(x,t|0,0)$ in Eq. \eqref{relax_et_1} will have different contributions depending on where $w^*$ (or equivalently $y$) lies. Finally, using Eq. \eqref{relax_et__871} in the formula of $P_r(x,t|0,0)$ in Eq. \eqref{relax_et_3}, we get 
\begin{align}
P_{r}(x,t) \sim e^{-t~ \mathcal{I}(y)}, ~~~~~~~~~\text{where}~y=\frac{|x|}{t^2},
\label{relax_et_4}
\end{align}
where $\mathcal{I}(y) = \text{min}\left[ \Phi(w,y)\right]$ and is given by 
\begin{align}
\mathcal{I}(y) &=\sqrt{\frac{8}{3} r^{3/2} y},~~~~~~~~~~\text{if~} y<\frac{2}{3}\sqrt{r}, \nonumber \\
&=r+\frac{3}{4}y^2, ~~~~~~~~~~~~\text{if~} y>\frac{2}{3}\sqrt{r}.
\label{relax_et_5}
\end{align}
We observe that the relaxation property of distribution has two different interpretations depending on the regime of $x$ . In the first regime $|x| <\frac{2}{3}\sqrt{r} t^2$, the system has already reached the stationary state. In fact the form of the distribution obtained in Eq. \eqref{relax_et_5} for this regime is consistent with the asymptotic results of $P_r^{st}(x)$ previously obtained in Eq. \eqref{reset_et_4}. On the other hand, the second regime $|x| >\frac{2}{3}\sqrt{r} t^2$ is still in the transient state. In Figure \ref{relax-et-1}, we have compared our temporal relaxation results with the numerical simulation. We have plotted both the stationary state and the large deviation form obtained above for three different times and compared them with the distribution obtained via numerical simulation. One clearly observes two distict regimes separated by $|x^*|$ and an excellent agreement with our analytical results. The front separating stationary state regime with the transient regime evolves super-ballistically with time as $x^*= \frac{2}{3}\sqrt{r} t^2$. We note that for Brownian particle, it was found in \cite{Majumdar2015} that the front moves ballistically.
\begin{figure}[t]
\includegraphics[scale=0.3]{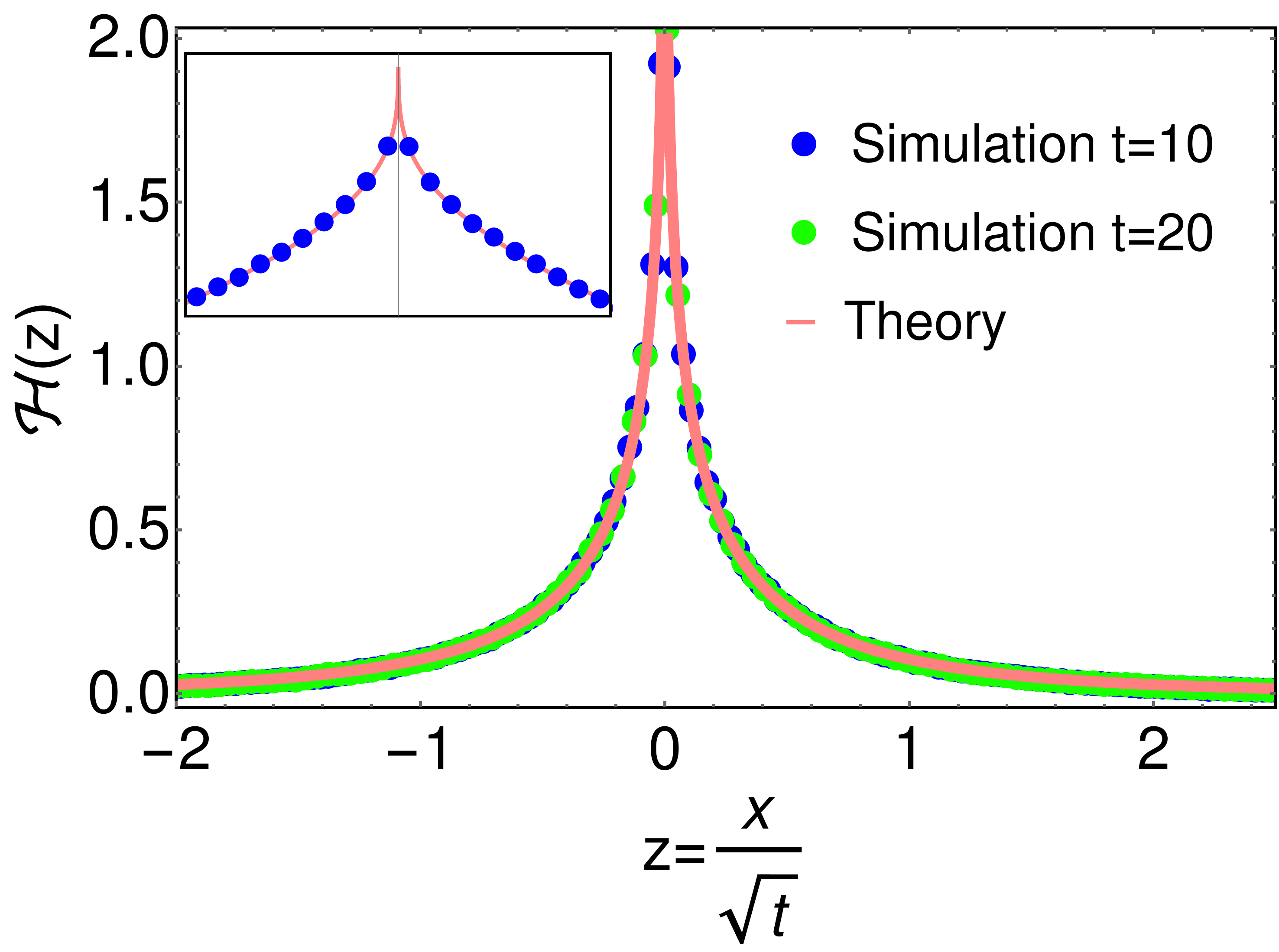}
\centering
\caption{Plot of the scaling function $\mathcal{H}(z)$ obtained in Eq. \eqref{scal-fun-eq-24} and comparison with the numerical simulation. We have conducted the comparison for $t=10$ and $t=20$ with $r=2$ for both. \textit{Inset:} Shows the same plot in log scale.}    
\label{prob-eta-0}
\end{figure}

\subsection{When only $x$ is reset (partial resetting)}
\label{prob-x}
\begin{figure}[t]
\includegraphics[scale=0.3]{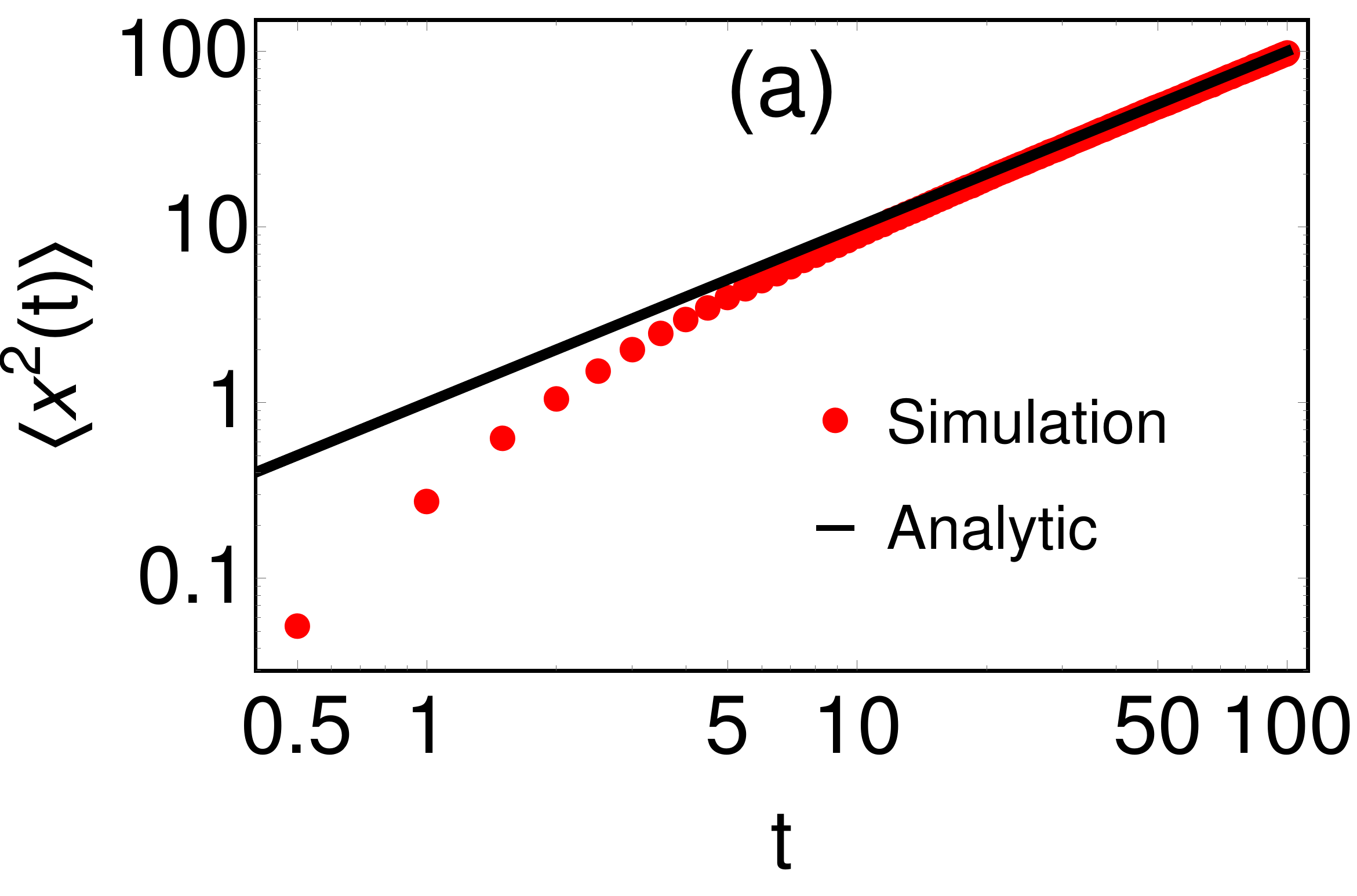}
\includegraphics[scale=0.3]{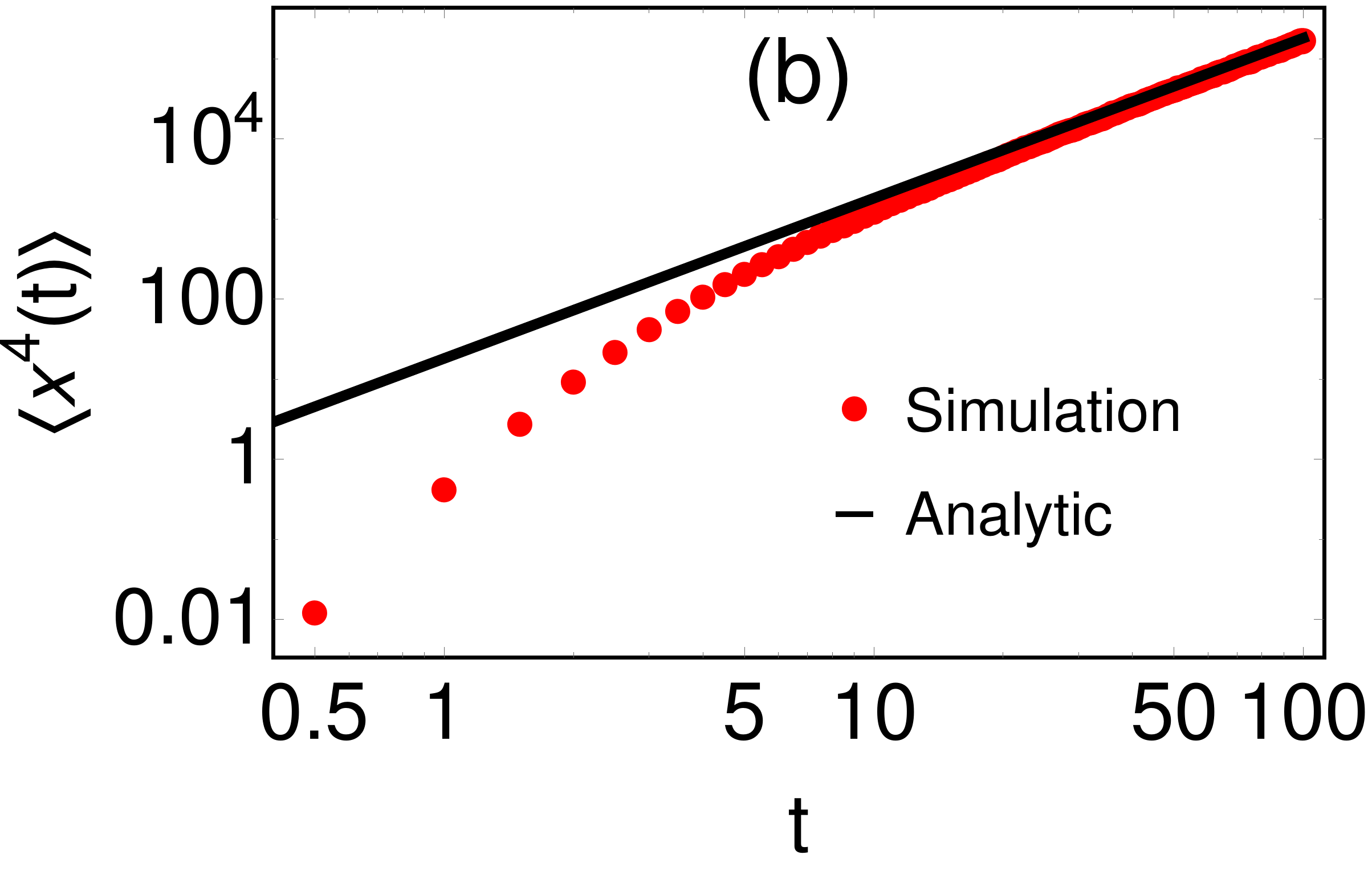}
\centering
\caption{In Figure (a), we plot the second moment of position $\langle x^2(t) \rangle$ (shown by black solid line) obtained in Eq. \eqref{reset-eta-0-eq-12} and compare it with the result of numerical simulations (shown by red filled circles). The match between them becomes better as we go to higher $t$. We have chosen $r=2$. In Figure (b), we have carried the same study for fourth moment $\langle x^4(t) \rangle$ for the same choice of parameters.}    
\label{moments-et-0}
\end{figure}

We now consider the case of partial resetting protocol. Here the dynamics is given by Eq. \eqref{protocol-eq-2}. For this protocol, the position $x$ is reset to its initial value $0$ at a constant rate $r$ while $v$ evolves without interruption. We denote the joint probability distribution to be at $(x,v)$ in time $t$ given that the particle initially was at $(0,0)$ as $\mathcal{P}_{r}(x,v,t|0,0)$ (which is different from $P_r(x,v,t|0,0)$ written for the complete resetting protocol). The last renewal equation reads,
 \begin{align}
\mathcal{P}_{r}(x,v,t|0,0)= e^{-r t} P_0(x,v,t|0,0)+r \int_{0}^{t} d\tau ~dv'~dx' e^{-r \tau} \mathcal{P}_{r}(x',v',t-\tau|0,0) P_0(x,v, \tau|0,v').
\label{reset-eta-0-eq-2}
\end{align}
The two terms appearing in the right hand side have same physical interpretation as that for the complete resetting case. The first term arises from those trajectories for which the particle has not experienced any resetting event upto time $t$. The second term arises from those trajectories for which the last reset happened at time $t-\tau$ when particle is at $(x',v')$. During this last reset event, the state of particle changes to $(0,v')$ which is in contrast to the complete resetting where the state changes to $(0,0)$. Starting from $(0,v')$ particle reaches $(x,v)$  in the time interval $t-\tau$ to $t$ during which it does not experience any reset event - the probability of which is $e^{-r \tau}$. It is worth emphasising that while integration over $x'$ can be carried out in the second term of $\mathcal{P}_{r}(x,v,t|0,0)$, integration over $v'$ cannot be performed. Therefore we cannot get rid of $\mathcal{P}_r$ term in contrast to the complete resetting where we were able to perform integration over $v'$. Finally to get distribution of $x$, we integrate $\mathcal{P}_{r}(x,v,t|0,0)$ over all $v$ to obtain,

\begin{align}
\mathcal{P}_{r}(x,t|0,0)= e^{-r t} P_0(x,t|0,0)+r \int_{0}^{t} d\tau ~dv'~ e^{-r \tau}~ p_{r}(v',t-\tau|0,0) P_0(x,\tau|0,v'),
\label{reset-eta-0-eq-1}
\end{align}
where,
\begin{align}
p_{r}(v,t|0,0) = \int_{-\infty}^{\infty} dx'~ \mathcal{P}_{r}(x',v,t|0,0).
\label{reset-erqe-123}
\end{align}
From the Langevin Eqs. \eqref{langevin-eq-2}, we see that the dynamics of $v$ is independent of $x$. Hence the resetting mechanism in $x$ does not affect the dynamics of $v$. This, in turn implies that $p_{r}(v,t|0,0)$ is just the probability distribution of $v$ in absence of resetting. Note that in absence of resetting, $v$ is simply Brownian motion. Hence we get, 
\begin{align}
p_{r}(v,t|0,0)=\frac{1}{\sqrt{4 \pi t}}e^{-\frac{v^2}{4 t}}.
\label{reset-eta-0-eq-2}
\end{align}
We now substitute $P_0(x,\tau|0,v')$ and $p_{r}(v',t-\tau|0,0)$ from  Eqs. \eqref{fokk_sol-eq2} and \eqref{reset-eta-0-eq-2} resepctively in $\mathcal{P}_{r}(x,t|0,0)$ in Eq. \eqref{reset-eta-0-eq-1}. Performing the integration over $v'$, we obtain

\begin{align}
\mathcal{P}_{r}(x,t|0,0)= e^{-r t} P_0(x,t|0,0)+r \int_{0}^{t} d\tau \frac{e^{-r \tau}}{\sqrt{4 \pi \sigma_{\tau}^2}} e^{-\frac{x^2}{4 \sigma_{\tau}^2}},
\label{reset-eta-0-eq-3}
\end{align}
where,
\begin{align}
\sigma_{\tau}&=\sqrt{\frac{\tau^2 (3 t-2 \tau)}{3}}.
\label{reset-eta-0-eq-4}
\end{align}
It is, in principle, possible to perform this integral for all $t$. We are, however, interested in the large $t$ behaviour of the distribution. For large $t$, the first term in $\mathcal{P}_{r}(x,t|0,0)$ decays exponentially fast. Hence only the second term will contribute for large $t$. 
\begin{align}
\mathcal{P}_{r}(x,t|0,0) \simeq r \int_{0}^{t} d\tau \frac{e^{-r \tau}}{\sqrt{4 \pi \sigma_{\tau}^2}} e^{-\frac{x^2}{4 \sigma_{\tau}^2}}.
\label{reset-eta-0-eq-333}
\end{align}
For large $t$, the integral will have bulk contribution from small $\tau$ limit. Replacing $\sigma _{\tau} \simeq \sqrt{\tau^2  t}$ , we can perfrom the integration over $\tau$. In  \ref{pr-v-0} , we have performed this integration for large $t$ to obtain the distribution  $\mathcal{P}_{r}(x,t|0,0)$. The final expression has the scaling structure, 

\begin{align}
\mathcal{P}_{r}(x,t|0,0)\simeq \frac{1}{\sqrt{t}} ~\mathcal{H}\left(\frac{|x|}{\sqrt{t}} \right),
\label{reset-eta-0-eq-51}
\end{align}
where the scaling function $\mathcal{H}(z)$ is given in terms of Meijner G-function $G$ \cite{Meijner} as,
\begin{align}
\mathcal{H}(z) = \frac{r}{4 \pi } \MeijerG{3}{0}{0}{3}{-}{0,0,\frac{1}{2}}{\frac{z^2 r^2}{16 }}.
\label{scal-fun-eq-24}
\end{align}
From the expression of $\mathcal{P}_{r}(x,t|0,0)$, we see that even though $x$ scales diffusively with $t$ (illustrated further in Figure \ref{moments-et-0}.(a)), however the scaling function $\mathcal{H}(z)$ is not Gaussian function. Instead, the scaling function is given by Eq. \eqref{scal-fun-eq-24} in terms of Meijner G-function. In Figure \ref{prob-eta-0}, we have plotted $\mathcal{H}(z)$ and compared with the numerical simulation for two different times. We observe an excellent match between them. Using the asymptotic forms of Meijner G-function, we get the following behaviour for scaling function,
\begin{align}
%\mathcal{H}(z)&\simeq \frac{r}{4 \sqrt{\pi }} \left[ \log \left( \frac{16 }{z^2 r^2}\right) - 2 \gamma _{E} +\psi_0(1/2)\right],
\mathcal{H}(z)&\simeq \frac{r}{4 \sqrt{\pi }} \left[ \log \left( \frac{16 }{z^2 r^2}\right) - 2 \gamma _{E} +\psi_0(1/2)\right],~~~\text{as }z \to 0,\\
& \simeq \frac{(2 r)^{2/3}}{\sqrt{3}z^{1/3}} ~\text{exp}\left(-\frac{3 r^{2/3} z^{2/3}}{2^{4/3}}\right), ~~~~~~~~~~~~~~~~\text{as }z \to \infty,
\label{reset-eta-0-eq-7}
\end{align}
where $\gamma _E$ is the Euler's constant and $\psi $ stands for PolyGamma function. From the asymptotic form we see that $\mathcal{H}(z)$ diverges logarithmically as $|z| \to 0$. This means that the distribution function $P_r(x,t|0,0)$ diverges as $\log (|x|)$ when $|x| \to 0$. Recall that the stationary state for complete resetting diverges as $\sim |x|^{-1/3}$ as $|x| \to 0$ as shown in Eq. \eqref{reset_et_4}. Moreover from the scaling structure of the distribution in Eq. \eqref{reset-eta-0-eq-51}, we see that the distribution does not have stationary state. 

It is worth remarking that for the case of underdamped Brownian motion (in which $v$ evolves via Eq. \eqref{langevin-eq-2} but with an additional damping term) the particle still exhibits stationary state for partial resetting as shown in \cite{Gupta2019}. Physically this can be understood in the following way. For RAP, the $v$ variable in absence of resetting performs the overdamped Brownian motion. The typical fluctuations of $v$ grows with time as $v \sim \sqrt{t}$. Even if the particle is brought to the origin at some rate, it can move far away from the origin at the next instant as velocity typically can be very large. On the other hand, $v$ for underdamped Brownian motion is bounded due to the presence of damping term. Therefore after some time, the typical flutuations of $v$ does not change further. When the particle is brought to the origin at some rate $r$, the extent over which the particle is localised does not change with time. This will result in the stationary state. To check the consistency of our results with \cite{Gupta2019}, one can take the limit of vanishing damping in \cite{Gupta2019} (see Eq. (36) in \cite{Gupta2019}) and find that the stationary state indeed vanishes in this limit.\\
After analysing the probability distribution, we now study the moments of $x$. By $x \to -x$ symmetry of the distribution, one immediately anticipates that the odd moments must vanish. To obtain the even moments, we take the Fourier transform of $\mathcal{P}_r(x,t|0,0)$ with respect to $x$ as $\mathcal{\bar{P}}_{r}(k,t)=\int_{-\infty}^{\infty} d x ~e^{i k x} \mathcal{P}_r(x,t|0,0)$. Inserting the expression of $\mathcal{P}_r(x,t|0,0)$ for large $t$ from Eq. \eqref{der-pr-v-0-eq-4} and performing the integration over $x$, we get

\begin{align}
\mathcal{\bar{P}}_r(k,t)&\simeq r t \int_{0}^{1} d w~ e^{-r t w-\bar{k}^{2} w^2}, \nonumber \\
&\simeq \frac{r t \sqrt{\pi}}{2 \bar{k}}e^{\frac{r^2 t^2}{4\bar{k}^2}} \left[ \text{Erfc}\left( \frac{r t}{2 \bar{k}}\right)-\text{Erfc}\left( \bar{k}+\frac{r t}{2 \bar{k}}\right)\right]
\label{reset-eta-0-eq-10}
\end{align}
where $\bar{k}=k~ t^{3/2}$. Note that 
\begin{align}
\langle x^m(t) \rangle =\frac{1}{i^m}\left[ \frac{d^m}{d k^m} \mathcal{\bar{P}}_r(k,t)\right] _{k \to 0}
\label{reset-eta-0-eq-11}
\end{align}
 We use the series representation of Erfc function as $\text{Erfc}\left(\frac{a}{k} \right)=\frac{k e^{-\frac{a^2}{k^2}}}{\sqrt{\pi} a} \sum_{n=0}^{\infty} (-1)^n\frac{(2n-1)!!}{2^n} (\frac{k}{a})^{2n}$, where $(2n-1)!!$ is product of all odd positive integers till $(2n-1)$ and is $1$ when $n=0$ . Using this representation, one finds that the second term inside the square bracket in Eq \eqref{reset-eta-0-eq-10} contibutes terms of the order $\sim e^{-r t}$ which can be neglected for large $t$. The leading contribution then comes only from first term which  finally gives,
\begin{align}
\mathcal{\bar{P}}_r(k,t) \simeq \sum_{n=0}^{\infty} (-1)^n \frac{2^n (2n-1)!!}{(r )^{2n}} \left( k \sqrt{t}\right)^{2n}
\label{reset-eta-0-eq-111}
\end{align} 
Using Eq. \eqref{reset-eta-0-eq-11}, one obtains the moments as,

\begin{align}
\langle x^m(t) \rangle &\simeq\frac{\sqrt{2^m} m!~ (m-1)!!}{r^m} t^{m/2}~~ \text{if } m \text{ is even. } \nonumber \\
& =0 ~~~~~~~~~~~~~~~~~~~~~~~~~~~~~ \text{if } m \text{ is odd. }
\label{reset-eta-0-eq-12}
\end{align}
As discussed before, we obtain that the odd moments vanish. We emphasize that the approximate equality in Eq. \eqref{reset-eta-0-eq-12} indicates that the expressions are valid only in large $t$. We have plotted the second and fourth moments in Figure \ref{moments-et-0} and compared them with numerical simulations. We see an excellent match between them.

\begin{figure}[t]
\includegraphics[scale=0.25]{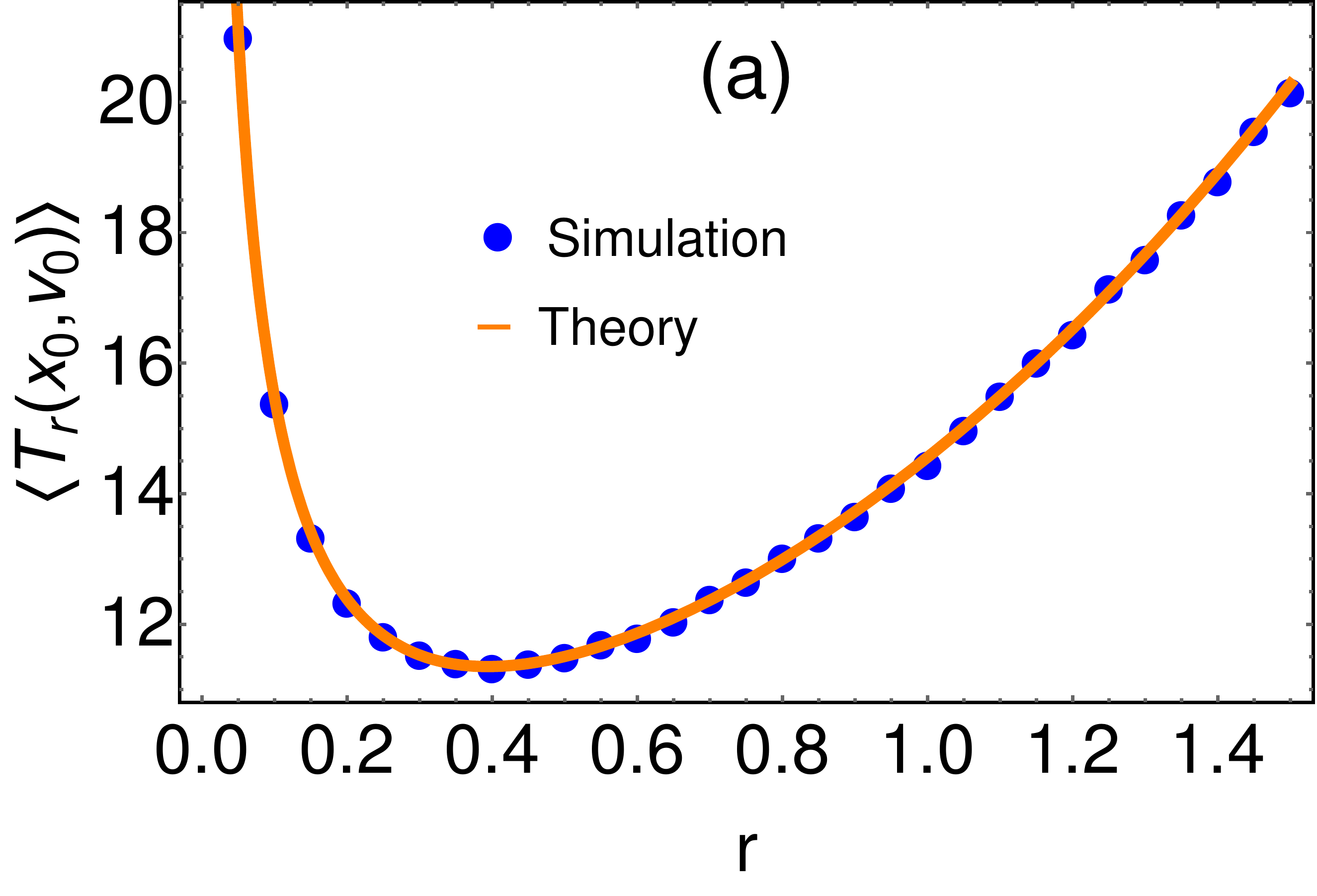}
\includegraphics[scale=0.25]{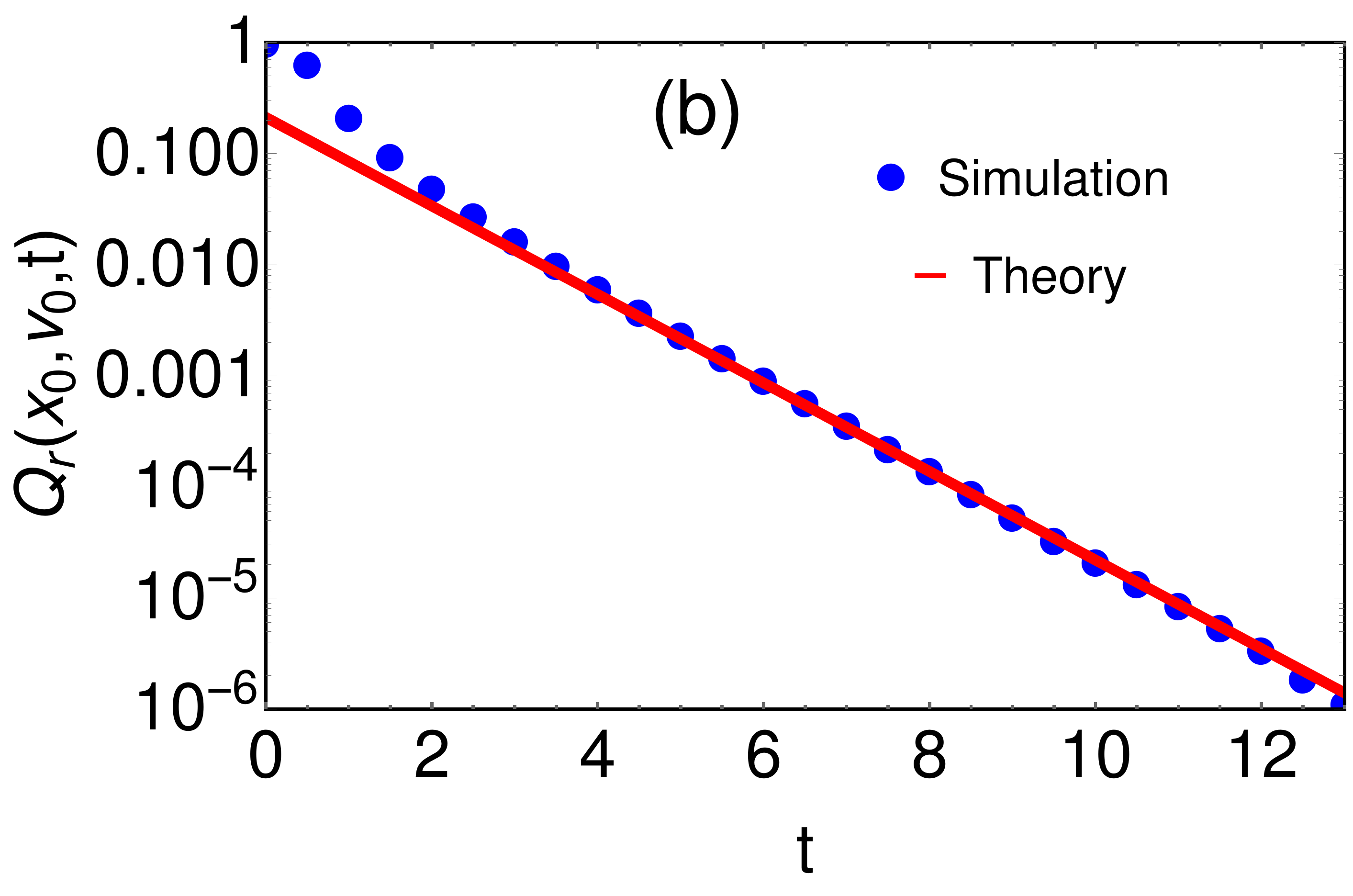}
\centering
\caption{(a) Comparision of $\langle T_r(x_0,v_0) \rangle$ from Eq. \eqref{surv-eta-1-eq-7} with the results of numerical simulation for resetting protocol 1. We have taken $x_0=1~v_0=0.5$. For this choice of parameters, the minima in $\langle T_r(x_0,v_0) \rangle$ occurs at $r^* = 0.377$. (b) Survival probability $Q_r(x_0,v_0,t)$ in Eq. \eqref{surv-eta-1-eq-8} is plotted and compared with the same from simulation. We see that for small $t$, theoretical and simulation results do not match. However as we move to large $t$ the agreement becomes better. The comparision is done for $x_0=1~v_0=-2$ and $r=1$. For this choice, $s_0 = -0.917324$ and $\mathcal{B} = 0.2111$.}    
\label{fig-surv-eta-1}
\end{figure}

\section{Survival probability}
\label{survvival}
First passage properties of stochastic processes under resetting have been quite extensively studied in recent time \cite{EvansSatya2011,Evans2011,EvansSatya2013,PalShlomi2017,PalChaterjee2019,EvansMajumdar2018,Masoliver2019,Kusmierz2014,KusmierzNowak2015}. 
Although most of these works have remained, so far, theoretical, there have been some recent experimental developments in the field of resetting \cite{Friedman2020, Besga2020}. Again, the paradigm of these studies is a diffusing particle subject to resetting mechanism. It is now well known that the mean first passage time (henceforth MFPT) of such a diffusing particle can be made finite by carefully using a resetting mechanism \cite{EvansSatya2011, EvansSatya2013, PalShlomi2017, EvansMajumdar2018}. More surprisingly, it is observed that resetting does not only lower the MFPT, it often renders it to the minimum hinting towards the emergence of an optimal resetting rate. This has been studied in great detail in \cite{PalShlomi2017, PrasadPal2019}. In what follows, we study the first passage propertied of RAP under the effect of resetting. We begin by reviewing the first passage properties of RAP in absence of resetting .  
\subsection{Survival probability in absence of resetting}
\label{surv-no-reset}
In this section, we will briefly review some known results for the first passage properties of RAP for brevity. We refer to \cite{Burkhardt2016} for a more detailed review. Let us denote $\mathcal{Z}(x,v,t|x_0,v_0)$ as the propagator to reach $(x,v)$ in time $t$ starting from $(x_0,v_0)$ in presence of an absorbing barrier at $x=0$. We will assume $x_0 \geq 0$. Given $\mathcal{Z}(x,v,t|x_0,v_0)$, it is possible to compute the survival probability $Q_0(x_0,v_0,t)$ by integrating it over all possible $v$ and $x$.
\begin{align}
Q_0(x_0,v_0,t) = \int _{-\infty}^{\infty} dv \int_{0}^{\infty} dx ~\mathcal{Z}(x,v,t|x_0,v_0).
\label{no-reset-surv-eq-1}
\end{align}
The forward master equation for $\mathcal{Z}(x,v,t|x_0,v_0)$ read as,
\begin{align}
\frac{\partial \mathcal{Z}}{\partial t}=\frac{\partial^2 \mathcal{Z}}{\partial v^2}-v \frac{\partial \mathcal{Z}}{\partial x}.
\label{no-reset-surv-eq-2}
\end{align} 
We have to solve this equation with appropriate boundary conditions. The first condition is $\mathcal{Z}(x,v,t|x_0,v_0)$ as $x \to \infty$ and/or $v \to \pm \infty$ is finite for all $t$. The other condition, due to the absorbtion barrier at $x=0$, is given by,
\begin{align}
\mathcal{Z}(x=0,v,t|x_0,v_0) = 0, ~~~~v>0.
\label{no-reset-surv-eq-3}
\end{align} 
We take the Laplace transform of the propagator with respect to $t$ as $\mathcal{\bar{Z}}(x,v,s|x_0,v_0)=\int_0^{\infty} dt e^{-s t}\mathcal{Z}(x,v,t|x_0,v_0)$. It was shown in \cite{Burkhawdt1993} that the Laplace transform is given by,
\begin{align}
\mathcal{\bar{Z}}(x,v,s|x_0,v_0) = \bar{P}_0(x,v,s|x_0, v_0) + \mathcal{\bar{Z}}_1(x,v,s|x_0, v_0),
\label{no-reset-surv-eq-4}
\end{align}
where $\bar{P}_0(x,v,s|x_0, v_0)$ is given by Eq. \eqref{fokk_sol1} and $\mathcal{\bar{Z}}_1(x,v,s|x_0, v_0)$ is given by,
\begin{align}
\mathcal{\bar{Z}}_1(x,v,s|x_0,v_0)& = -\frac{1}{2 \pi} \int_{0}^{\infty}\frac{dF}{F^{1/6}} \int_{0}^{\infty}\frac{dG}{(F+G)G^{1/6}} \text{exp} \left[ -F x-G x_0 -\frac{2 s^{3/2}}{3} \left(\frac{1}{F}+\frac{1}{G} \right)\right] \nonumber \\
&~~~~~~~~~~~~~~~\text{Ai} \left[ -v F^{1/3}+\frac{s}{F^{2/3}}\right]~\text{Ai} \left[ v_0 G^{1/3}+\frac{s}{G^{2/3}}\right].
\label{no-reset-surv-eq-5}
\end{align}
Denoting the Laplace transform of $Q_0(x_0,v_0,t)$ in $t$ as $\bar{Q}_0(x_0,v_0,s)$, we integrate $\mathcal{\bar{Z}}_1(x,v,s|x_0,v_0)$ over all possible $x$ and $v$ as shown in  Eq. \eqref{no-reset-surv-eq-1} to get $\bar{Q}_0(x_0,v_0,s)$. Performing this integral \cite{Burkhawdt1993,Burkhardt2016} gives,
\begin{align}
\bar{Q}_0(x_0,v_0,s)=\frac{1}{s}-\alpha(s),
\label{surv-eta-1-eq-1}
\end{align}
where 
\begin{align}
\alpha (s)=\int_0^{\infty}\frac{dy}{y^{5/3}} e^{-y x_0} \text{Ai}\left( v_0 y^{1/3}+\frac{s}{y^{2/3}} \right) \left[ 1+\frac{1}{4 \sqrt{\pi}} \Gamma \left( -\frac{1}{2},\frac{2 s^{3/2}}{3 y} \right)\right],
\label{surv-eta-1-eq-2}
\end{align}
where $\Gamma (\beta, z)$ is the incomplete gamma function. Inverting Eq. \eqref{surv-eta-1-eq-2} for arbitrary $s$ turns out to be very difficult. However for small $s$ (which implies large $t$) one can get an approximate expression for the survival probability as given in \cite{Burkhawdt1993}.
\begin{align}
&\bar{Q}_0(x_0,v_0,s) \simeq \frac{3^{5/6} \sqrt{2 \pi} x_0^{1/6}}{s^{3/4}} H\left(-v_0 x_0^{-1/3}\right),~~~~~\text{with}\label{surv-eta-1-eqqq-202}\\
& H(y) = \frac{U\left( -\frac{1}{6},\frac{2}{3},-\frac{y^3}{9}\right)}{\Gamma\left( \frac{1}{6}\right) \Gamma\left( \frac{2}{3}\right)} + \frac{y}{9^{\frac{1}{3}}} \frac{U\left( \frac{1}{6},\frac{4}{3},-\frac{y^3}{9}\right)}{\Gamma\left(- \frac{1}{6}\right) \Gamma\left( \frac{4}{3}\right)},
\label{H-def}
\end{align} 
where $U$ stands for the Kummers confluent hypergeometric function. Inverting the expression in Eq. \eqref{surv-eta-1-eqqq-202} to get the survival probability for large $t$ is now straight forward. The inversion of $\bar{Q}_0(x_0,v_0,s)\sim \frac{1}{s^{3/4}}$ gives $Q_0(x_0,v_0,t)\sim \frac{1}{t^{1/4}}$ for RAP \cite{Burkhawdt1993,Burkhardt2016}. We close this section by remarking that the same expression for $Q_0(x_0,v_0,t)$ for large $t$ was also obtained in \cite{ScherMajumdar2013} by solving the backward Fokker-Planck equation for $Q_0(x_0,v_0,t)$.

\section{Survival probability in presence of resetting}
\label{surv-reset}
\subsection{When both $x$ and $v$ are reset (complete resetting)}
\label{survv-x-v}
We now consider the first passage properties of RAP when both $x$ and $v$ are reset at a rate $r$ to their initial values $x_0$ and $v_0$ respectively (see Eq. \eqref{protocol-eq-1}). Denoting the survival probability by $Q_r(x_0,v_0,t)$, we write the last renewal equation as,
\begin{align}
Q_{r}(x_0,v_0,t)=e^{-r t} Q_{0}(x_0,v_0,t)+r \int_{0}^{t} d \tau e^{-r \tau} Q_{0}(x_0,v_0, \tau) Q_{r}(x_0,v_0, t-\tau).
\label{surv-eta-1-eq-3}
\end{align}
The first term in the R.H.S. arises from those realisations for which the particle has neither reset nor reached the origin till time $t$. The second term integrates the contribution of those trajectories for which the last reset happens at time $t-\tau$ upto which the particle also survives the absorbtion wall at the origin. After the last reset, the state of particle changes to $(x_0,v_0)$. Starting from this state, the particle does not further reset in the remaining interval $t-\tau$ to $t$ and also survives the absorbtion barrier.

Proceeding further, we solve the renewal Eq. \eqref{surv-eta-1-eq-3}. Taking the Laplace transform of $Q_{r}(x_0,v_0,t)$ in $t$,
\begin{align}
\bar{Q}_r(x_0,v_0,s) = \int_0^{\infty} dt~ e^{-s t}~ Q_{r}(x_0,v_0,t),
\label{lap-eq-rw-1}
\end{align}
the renewal Eq. \eqref{surv-eta-1-eq-3} can be recasted in terms of the Laplace transforms as,
\begin{align}
&\bar{Q}_{r}(x_0,v_0,s)=\bar{Q}_{0}(x_0,v_0,s+r)+r \bar{Q}_{0}(x_0,v_0, s+r) \bar{Q}_{r}(x_0,v_0, s), \nonumber \\
&\bar{Q}_{r}(x_0,v_0,s)=\frac{\bar{Q}_{0}(x_0,v_0,s+r)}{1-r \bar{Q}_{0}(x_0,v_0,s+r)}.
\label{surv-eta-1-eq-4}
\end{align}
We have expressed $\bar{Q}_{r}(x_0,v_0,s)$ in terms of $\bar{Q}_0(x_0,v_0,r+s)$ which is given by Eq. \eqref{surv-eta-1-eq-1}. This formula is quite important since it connects process with resetting to the process without resetting. In other words, we can use the properties of the underlying dynamics (namely Eq. \eqref{surv-eta-1-eq-1}) to predict the effects of resetting. To get the survival probability in the time domain, we need to invert the Laplace transform $\bar{Q}_{r}(x_0,v_0,s)$ in Eq. \eqref{surv-eta-1-eq-4}. However before inverting, we look at the MFPT in presence of resetting. \\
\textit{Mean first passage time:} The MFPT $\langle T_r(x_0,v_0) \rangle$ is defined in terms of survival probability as,
\begin{align}
\langle T_r(x_0,v_0) \rangle = -\int_0^{\infty} ~dt~t \frac{\partial Q_r(x_0,v_0,t)}{\partial t} =\bar{Q}_r(x_0,v_0,s \to 0).
\label{MFPTeqqq-q1}
\end{align}
Using the expression of $\bar{Q}_r(x_0,v_0,s \to 0)$ from \eqref{surv-eta-1-eq-4}, we have
\begin{align}
\langle T_r(x_0,v_0) \rangle &=\frac{\bar{Q}_{0}(x_0,v_0,r)}{1-r \bar{Q}_{0}(x_0,v_0,r)}, \label{surv-eta-1-eq-5} \\
&= -\frac{1}{r}+\frac{1}{r^2 \alpha (r)}.
 \label{surv-eta-1-eq-6}
\end{align}
While going to second line, we have used the expression of $Q_0(x_,v_0,s)$ from Eq. \eqref{surv-eta-1-eq-1} with $\alpha (r)$ given in Eq. \eqref{surv-eta-1-eq-2}. In Figure \ref{fig-surv-eta-1}(a), we have compared $\langle T_r(x_0,v_0) \rangle$ in Eq. \eqref{surv-eta-1-eq-6} with the numerical simulation. We observe an excellent agreement between them. From the figure we observe that $\langle T_r(x_0,v_0) \rangle$ exihibits minimum at some $r = r^*$. To explain the appearance of such behaviour for RAP, we study the asymptotic behaviours of $\langle T_r(x_0,v_0) \rangle$ with respect to $r$. For $r \to 0$, we see from Eq. \eqref{surv-eta-1-eq-5} that $\langle T_r(x_0,v_0) \rangle \simeq \bar{Q}_{0}(x_0,v_0,r)$ which from Eq. \eqref{surv-eta-1-eqqq-202} diverges as $\sim r^{-3/4}$. On the other hand for $r \to \infty$ ($r >>x_0^{-2/3}$ and $r>>v_0^2$), we have shown in \ref{MFPT-large-r} that $\langle T_r(x_0,v_0) \rangle \sim e^{\sqrt{\frac{8}{3}x_0 r^{3/2}}}$. The approximate expressions for $\langle T_r(x_0,v_0) \rangle$ in different limits of $r$ can be summarized as,

\begin{align}
\langle T_r(x_0,v_0) \rangle & \simeq \frac{3^{5/6} \sqrt{2 \pi} x_0^{1/6}}{r^{3/4}} H\left(-v_0 x_0^{-1/3}\right),~~~~\text{as }r \to 0, \nonumber \\
&\simeq \sqrt{\frac{8}{3 r^2}} e^{v_0 \sqrt{r}} e^{\sqrt{\frac{8}{3}x_0 r^{3/2}}},~~~~~~~~~~~~~\text{as }r \to \infty,
\label{surv-eta-1-eq-7}
\end{align}
where $H(y)$ is given by Eq. \eqref{H-def}. From the asymptotic form at $r \to 0$, we observe that increasing $r$ decreases $\langle T_r(x_0,v_0) \rangle$. On the other hand for $r \to \infty$, $\langle T_r(x_0,v_0) \rangle$ increases with increase in $r$. Therefore at some value $r$ (say $r^*$), $\langle T_r(x_0,v_0) \rangle$ will change the slope from negative to positive. The minimum of $\langle T_r(x_0,v_0) \rangle$ occurs at $r = r^*$ which is given by,
\begin{align}
\left[\frac{\partial }{\partial r} \langle T_r(x_0,v_0) \rangle \right]_{r=r^*} =0,
\label{min-value-T}
\end{align}
where $\langle T_r(x_0,v_0) \rangle$ is given in Eq. \eqref{surv-eta-1-eq-6}. Before clsoing this section, we remark that MFPT for other stochastic processes like Brownian motion has also been studied \cite{EvansSatya2011, EvansSatya2013, PalShlomi2017, EvansMajumdar2018}. For such cases also, MFPT can be optimised with respect to the resetting rate $r$. In this paper, we have extended these results for the randomly accelerated particle.\\
\textit{Survival probability:} Let us now try to invert $\bar{Q}_{r}(x_0,v_0,s)$ in Eq. \eqref{surv-eta-1-eq-4} to get the survival probability in time domain. Although one can, in principle, invert $\bar{Q}_{r}(x_0,v_0,s)$, we are interested in the large $t$ behaviour of $Q_r(x_0,v_0,t)$. Looking at the expression of $\bar{Q}_{r}(x_0,v_0,s)$ in Eq. \eqref{surv-eta-1-eq-4}, we find that it has pole at $s=s_0$ which is obtained by solving  $1-r \bar{Q}_0(x_0, v_0, r+s_0)=0$. The pole $s_0$ will set the time scale of decay for $Q_r(x_0,v_0,t)$ at large $t$. We obtain
\begin{align}
Q_{r}(x_0,v_0, t) \simeq \mathcal{B}~ \text{exp}(s_0 t),
\label{surv-eta-1-eq-8}
\end{align}
where approximate equality indicates that the relation is valid only in large $t$ limit. Also here, $s_0$ and $\mathcal{B}$ are given by,
\begin{align}
& 1-r \bar{Q}_0(x_0, v_0, r+s_0)=0, \label{surv-eta-1-eq-9}\\
& \mathcal{B}=-\frac{1}{r} ~\frac{\bar{Q}_0(x_0,v_0, r+s_0)}{\bar{Q}_0'(x_0,v_0,r+s_0)},
\label{surv-eta-1-eq-10}
\end{align}
where  $\bar{Q}_0'(x_0,v_0,s) = \frac{\partial}{\partial s}\bar{Q}_0(x_0,v_0,s)$. In figure \ref{fig-surv-eta-1}(b), we have plotted $Q_{r}(x_0,v_0,t)$ in Eq. \eqref{surv-eta-1-eq-8} and compared with the numerical simulation. We observe better match as we go to large $t$, the agreement becomes better. One can simplify the expressions in Eqs. \eqref{surv-eta-1-eq-9} and \eqref{surv-eta-1-eq-10} for $r x_0^{2/3}>>1$ and $v_0^2 r^{-1}<<1$. In these limits, the expression of $Q_r(x_0,v_0,r)$ as written in Eq. \eqref{surv-eta-1-eq-1} can be simplified. In these limits, the expression of $\alpha (r)$  is derived in \ref{MFPT-large-r} (see Eq. \eqref{MFPT-large-gam-eq-4}). Inserting this expression in Eq. \eqref{surv-eta-1-eq-1} to get $\bar{Q}_{0}(x_0,v_0,r)$ which can be used in Eq. \eqref{surv-eta-1-eq-9} to get
\begin{align}
\frac{s_0}{r+s_0} = -r \alpha (r+s_0).
\label{surv-eta-1-eq-1211}
\end{align}
We assume that $s_0 <<r$ and using $\alpha (r)$ from Eq. \eqref{MFPT-large-gam-eq-4}, we get
\begin{align}
&s_0\simeq-\sqrt{\frac{3}{8}}r ~e^{-v_0\sqrt{r}-\sqrt{
\frac{8}{3}r^{3/2} x_0}}, \label{surv-eta-1-eq-11} \\
&\mathcal{B}\simeq 1.
\label{surv-eta-1-eq-12}
\end{align}
Substituting this in Eq. \eqref{surv-eta-1-eq-8}, we get for $r x_0^{2/3}>>1$ and $v_0^2 r^{-1}<<1$,
\begin{align}
Q_{r}(x_0,v_0, t) \simeq \text{exp}\left({-\sqrt{\frac{3}{8}} rt~ e^{-\gamma}}\right),
\label{surv-eta-1-eq-13}
\end{align}
where $\gamma =v_0 \sqrt{r}+\sqrt{\frac{8}{3} r^{3/2}x_0}~$. We obtain that $Q_r(x_0,v_0,t)$ decays exponentially with a time scale $\tau _r = \left[ \sqrt{\frac{3}{8}} r~ e^{-\gamma}\right]^{-1}$. We see that Eq. \eqref{surv-eta-1-eq-13} has the form of Gumble distribution which appears widely in the extreme value statistics \cite{extreme1,extreme2}. The appearance of  the Gumble distribution in the survival probability was also seen for Brownian motion in the seminal work by Evans and Majumdar in \cite{Evans2011}.

\begin{figure}[t]
\includegraphics[scale=0.26]{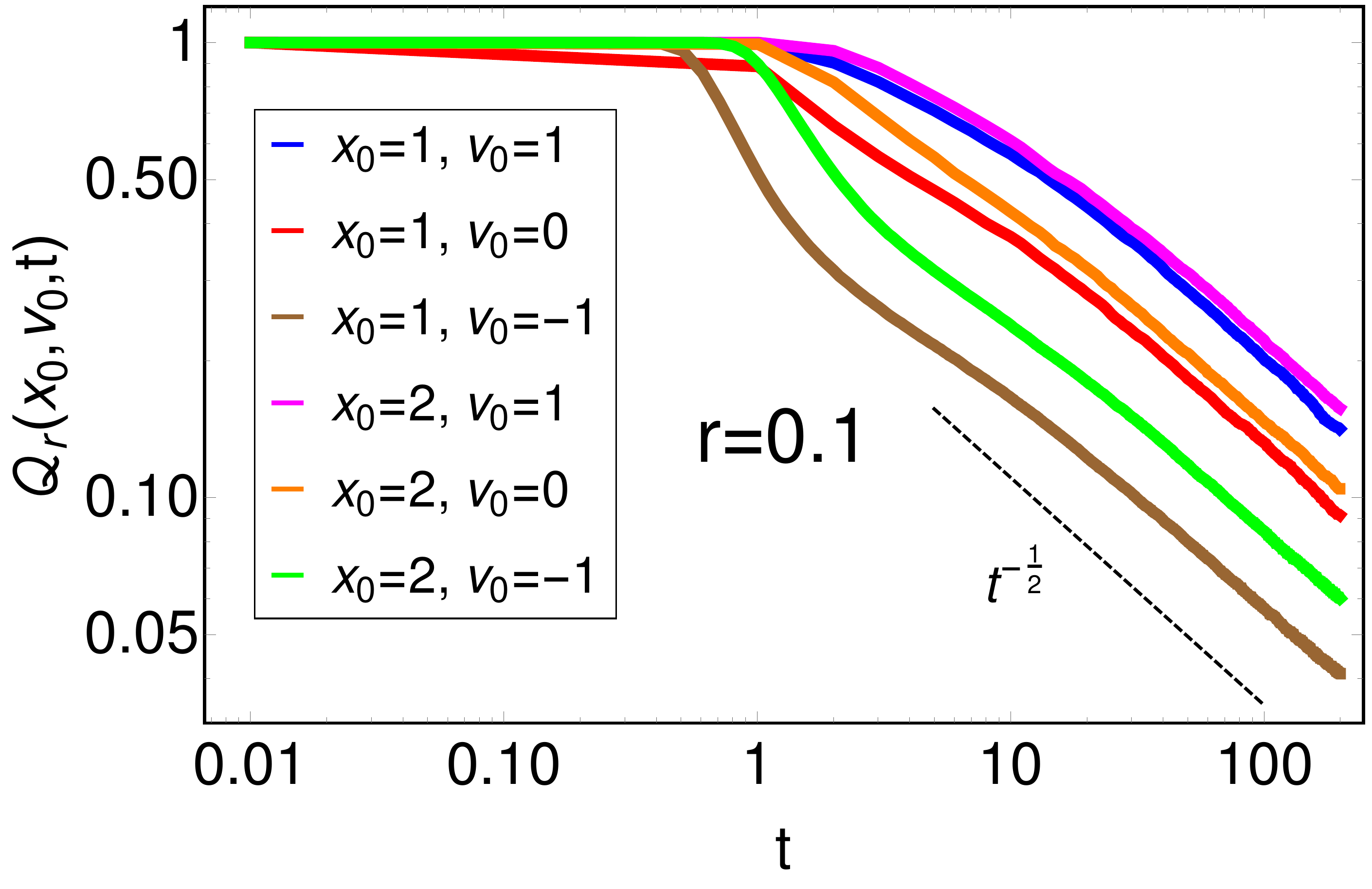}
\includegraphics[scale=0.26]{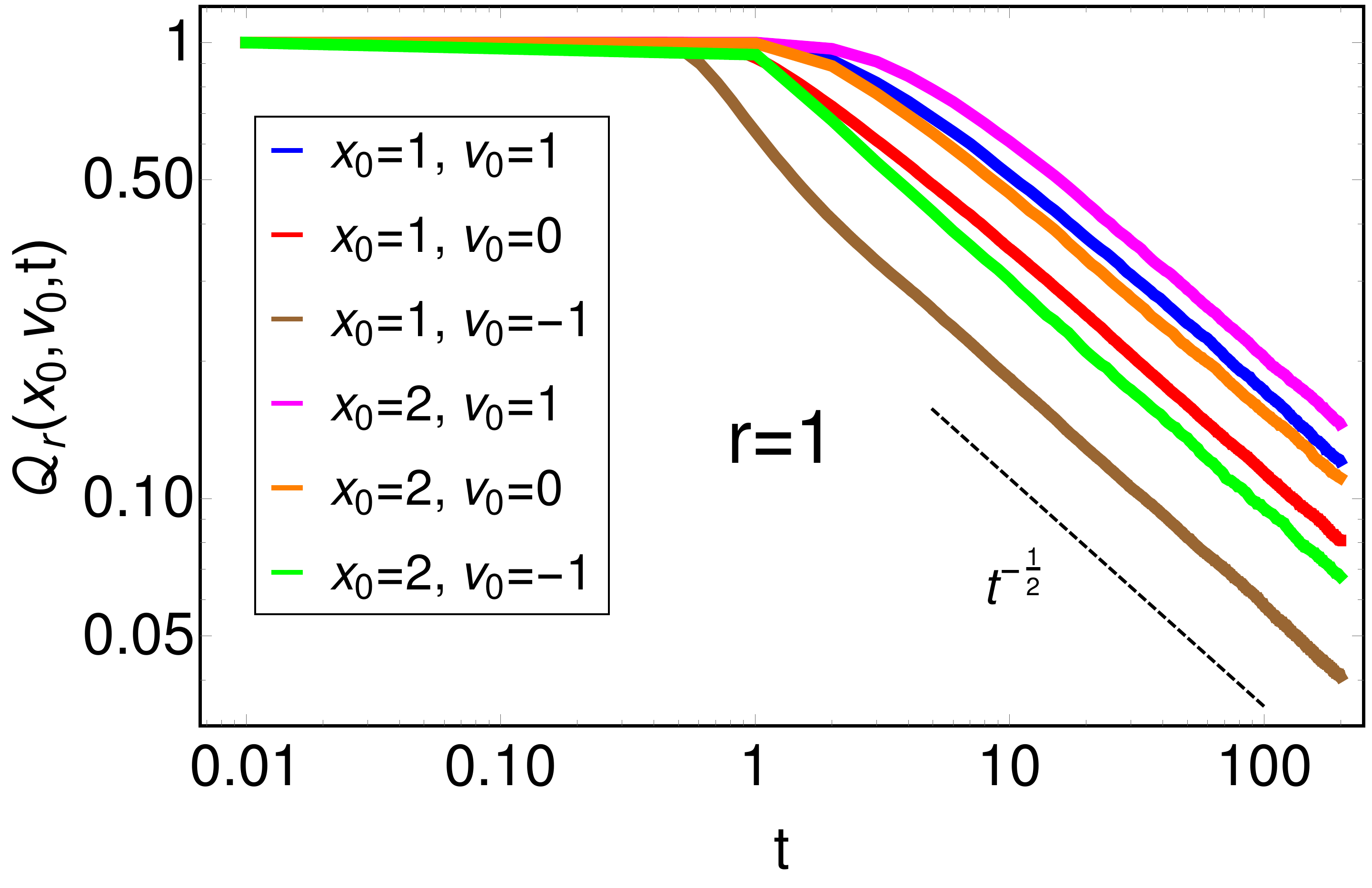}
\centering
\caption{Numerical simulation of the survival probability $\mathcal{Q}_r(x_0,v_0,t)$ for partial resetting protocol for different values of $x_0$ and $v_0$. In the left panel, we have shown the results for $r=0.1$. The black dotted line is a plot of $\frac{1}{\sqrt{t}}$. We observe that $\mathcal{Q}_r(x_0,v_0,t)$ decays as $\frac{\mathcal{A}}{\sqrt{t}}$ at large times for all values of $x_0$ and $v_0$. However the value of $\mathcal{A}(x_0,v_0)$ depends on $x_0$ and $v_0$. The right panel shows the same plot for $r =1$.}    
\label{fig-surv-eta}
\end{figure}

\subsection{When only $x$ is reset (partial resetting)}
\label{surv-x}
Let us next consider the first passage properties for the particle under partial resetting defined in Eq. \eqref{protocol-eq-2}. We denote the survival probability by $\mathcal{Q}_{r}(x_0,v_0,t)$ (which is different than $Q_{r}(x_0,v_0,t)$ for the complete resetting). We write the last renewal equation for this case as,
\begin{align}
\mathcal{Q}_r(x_0,v_0,t) = e^{-r t} Q_0(x_0,v_0,t) + r \int_{0}^{t} d \tau~dv~ e^{-r \tau} ~z_{r}(v,t-\tau|x_0,v_0) Q_{0}(x_0,v, \tau),
\label{ssurv-eta-0-eq-1}
\end{align}
with $z_r(v,t|x_0,v_0)$ defined in terms of the propagator $\mathcal{Z}_r(x,v,t|x_0,v_0)$ in presence of absorbing barrier at $x=0$ as,
\begin{align}
z_r(v,t|x_0,v_0) = \int_{0}^{\infty} dx~\mathcal{Z}_r(x,v,t|x_0,v_0).
\label{ssurv-eta-0-eq-2}
\end{align}
The two terms appearing on the R.H.S. of the renewal Eq. \eqref{ssurv-eta-0-eq-1} have same physical interpretation as that in the complete resetting case. The first term arises from those realisations for which the particle has neither reset nor reached the origin till time $t$. The second term arises from those realisations for which last reset happens at time $t-\tau$ and particle has survived the absorbtion wall. At this instant the state of the particle changes to $(x_0,v)$. Starting from this state, the particle does not further reset and also survives the absorbtion wall.

Contrary to the case of free propagator, we were not able to compute $z_r(v,t|x_0,v_0)$  in presence of absorbing boundary wall at $x=0$. However we present some interesting results based on the numerical study. In Figure \ref{fig-surv-eta}, we have plotted $\mathcal{Q}_r(x_0,v_0,t)$ vs $t$ obtained from numerical simulation for various values of $r$, $x_0$ and $v_0$. From the plot we observe that $\mathcal{Q}_r(x_0,v_0,t)$ for large $t$ decays as,
\begin{align}
\mathcal{Q}_r(x_0,v_0,t) \simeq \frac{\mathcal{A}(x_0,v_0)}{\sqrt{t}},
\label{ssurv-eta-0-eq-3}
\end{align}
where $\mathcal{A}$ is a constant which depends on $x_0,~v_0$ and $r$. Eq. \eqref{ssurv-eta-0-eq-3} implies that first-passage probability distribution decays as $\sim t^{-3/2}$ which means that the moments of first-passage time are still divergent. Therefore we observe that the partial resetting protocol does not render moments of first-passage time finite. Hence for search purposes, the complete resetting protocol is more efficient than the partial one. To understand this physically, note that the fluctuation in $v$ grows with time diffusively as $v \sim \sqrt{t}$. Under the partial resetting, even though the particle is brought to $x_0$ at some rate, it can move substantially far at the next instant since $v$ in unconstrained and can,in principle, be quite large. In other words, it can take inifinite time to hit the absorbing barrier at $x=0$. This physical intuition essentially asserts that the moments of first passage time, even in the presence of resetting, diverge.

\section{Conclusions}
\label{conclusion}
In this paper, we have studied a simple non-Markov model namely a randomly accelerated particle in one dimension under stochastic resetting mechanism. We considered two resetting protocols namely (i) complete resetting: both $x$ and $v$ reset to their initial values $x_0$ and $v_0$ at a constant rate $r$ and (ii) partial resetting: only $x$ resets to $x_0$ while $v$ evolves uninterruptedly. In first part of the paper, we analysed the model in an infinite line. For complete resetting, we showed that the particle reaches to a non-equilibrium steady state which is the fingerprint characteristics of the resetting systems. We also investigated the late time relaxation properties of the distribution for this case. The late time relaxation possessed qualitatvely same behaviour as seen for Brownian motion \cite{Majumdar2015}, which we extended for RAP in this paper. In particular, we showed the existence of a travelling front at $x^*=\frac{2}{3}\sqrt{r} t^2$ that separates the two regions of stationary state and transient state. While for $x <|x^*|$ stationary state has already been achieved, $x>|x^*|$ is still in the transient state . For the latter region, we showed that the distribution has large deviation form as shown in Eq.\eqref{relax_et_4} and the rate functions were computed in Eq. \eqref{relax_et_5}. Next, we studied the probability distribution for RAP under partial resetting. In stark contrast, here we showed that the particle is always in the transient state. We obtained that the position scales with time difusively as $x \sim \sqrt{t}$. The probability distibution for the typical fluctuations of $x$ follows a scaling form as shown in Eq. \eqref{reset-eta-0-eq-51}. We rigorously derived this scaling form as well as the corresponding scaling function in Eq. \eqref{scal-fun-eq-24} which is given in terms of Meijner G functions.\\
In the second part of the paper, we studied the first passage time properties of RAP under resetting. For complete resetting, we computed the mean first passage time and showed that it has an optimal value at some $r^*$ in Eq. \eqref{min-value-T}. We also obtained the late time behaviour of the survival probability which decays exponentially as $\sim \mathcal{B} e^{-|s_0|t}$ with $s_0$ and $\mathcal{B}$ given by Eqs. \eqref{surv-eta-1-eq-9} and  \eqref{surv-eta-1-eq-10} respectively. On the other hand, for partial resetting, we numerically showed that the survival probability decays at large times as $\mathcal{Q}_{r}\sim t^{-\frac{1}{2}}$. This means that the first passage time moments are infinite even in presence of resetting.\\
We note that in this paper we have focussed on two different resetting protocols. Studying other resetting protocols for this model is an interesting future direction. Recently it was shown that that the $y$- coordinate of the active Brownian particle in some approriate limit is described by the same Langevin equation as for RAP \cite{Basu2020a} . It would be interesting to study the behaviour of an active Brownian particle under the effect of resetting. In real world resetting is a physical process and takes finite amount of time. This scenario was considered and studied in detail for Brownian particle in \cite{Eva2018,Pal2019,PalKu12019,PalKu22019,Bodr2020, Mas2019,GuptaPlata2020}. Studying RAP under such resetting mechanisms is a promising direction for future studies.
\section{Acknowledgement}
I would like to thank my supervisor Dr. Anupam Kundu for his suggestions and encouragement throughout the work. I am also grateful to Arnab Pal for his critical comments about the work. I am also indebted to him  for meticulously reading the manuscript and providing many fruitful suggestions which finally led to this paper.

\appendix

\section{Derivation of $P_r^{st}(x)$ as $|x| \to \infty$ in Eq. \eqref{reset_et_4}}
\label{assy-x-pr}
In this appendix, we will derive the asymptotic expression of $P_r^{st}(x)$ for large $|x|$ as written Eq. \eqref{reset_et_4}. We consider the expression of $P_r^{st}(x)$ in Eq. \eqref{reset-et-300p}. Changing $F=w/|x|$ in this expression, we have
\begin{align}
P_r^{st}(x) = \frac{r}{|x|^{1/3}} \int_0^{\infty} \frac{dw}{w^{2/3}} e^{-w} \text{Ai} \left(\frac{r |x|^{2/3}}{w^{2/3}} \right).
\label{assy-x-pr-eq-1}
\end{align} 
For $r |x|^{2/3} \to \infty$, the argument of Airy function becomes very large for finite $w$. We use the asymptotic form of Airy function $\text{Ai}(z)\simeq \frac{\Gamma(1/6) \Gamma(5/6)}{4 \pi ^{3/2} z^{1/4}} e^{-\frac{2}{3}z^{3/2}}$ for $z \to \infty$. Inserting this in Eq. \eqref{assy-x-pr-eq-1}, one gets
\begin{align}
P_r^{st}(x) \simeq \frac{r^{3/4}\Gamma\left(\frac{1}{6} \right) \Gamma\left(\frac{5}{6} \right)}{4 \pi ^{3/2} \sqrt{|x|}} \int _0^{\infty} \frac{dw}{\sqrt{w}}\text{exp}\left( -w-\frac{2 r^{3/2} |x|}{3w} \right).
\label{assy-x-pr-eq-2}
\end{align}
Performing the integral in the right hand side we obtain the result quoted in Eq. \eqref{reset_et_4}.

\section{Derivation of $\mathcal{P}_r(x,t|0,0)$ in Eq. \eqref{reset-eta-0-eq-51} as $t \to \infty$}
\label{pr-v-0}
Here we derive the approximate expression of $\mathcal{P}_r(x,t|0,0)$ as given in Eq. \eqref{reset-eta-0-eq-51}. To begin with we rewrite Eq. \eqref{reset-eta-0-eq-333} as,
\begin{align}
\mathcal{P}_{r}(x,t|0,0) \simeq r \int_{0}^{t} d\tau \frac{e^{-r \tau}}{\sqrt{4 \pi \sigma_{\tau}^2}} e^{-\frac{x^2}{4 \sigma_{\tau}^2}}.
\label{der-pr-v-0-eq-1}
\end{align}
where $\sigma_{\tau}=\sqrt{\frac{\tau^2 (3 t-2 \tau)}{3}}$. Changing variable $\tau = t u$, we get
\begin{align}
\mathcal{P}_{r}(x,t|0,0) \simeq r t \int_{0}^{1} du~e^{-r t u}~\sqrt{\frac{3}{4 \pi t^3 (3u^2-2u^3)}}  e^{-\frac{3x^2}{4  t^3 (3u^2-2u^3)}}.
\label{der-pr-v-0-eq-2}
\end{align}
For $t \to \infty$, the integral will be dominated by $u \to 0$ due to the presence of $e^{-r t u}$ term. This means we can replace $3 u^2-2 u^3 \simeq 3 u^2$. Hence Eq. \eqref{der-pr-v-0-eq-2} can be simplified as,
\begin{align}
\mathcal{P}_{r}(x,t|0,0) \simeq r t \int_{0}^{1} du~e^{-r t u}~\sqrt{\frac{1}{4 \pi t^3 u^2}} e^{-\frac{x^2}{4  t^3 u^2}}.
\label{der-pr-v-0-eq-4}
\end{align}
Going back to $u =\tau/t$, we get
\begin{align}
\mathcal{P}_{r}(x,t|0,0) \simeq r  \int_{0}^{t} du~e^{-r \tau}~\sqrt{\frac{1}{4 \pi t \tau^2}} e^{-\frac{x^2}{4  t \tau^2}}.
\label{der-pr-v-0-eq-5}
\end{align}
Splitting the integral on right hand side as $\int _{0}^{t} = \int_{0}^{\infty} - \int_{t}^{\infty}$ and noting the $\int_{t}^{\infty}$ for $t \to \infty$ is very small, one gets
\begin{align}
\mathcal{P}_{r}(x,t|0,0) \simeq r  \int_{0}^{\infty} du~e^{-r \tau}~\sqrt{\frac{1}{4 \pi t \tau^2}} e^{-\frac{x^2}{4  t \tau^2}}.
\label{der-pr-v-0-eq-6}
\end{align}
Performing the intergration using \textit{Mathematica}, we obtain the result in Eq. \eqref{reset-eta-0-eq-51}.

\section{$\langle T_{r}(x_0,v_0) \rangle$ as $r \to \infty$}
\label{MFPT-large-r}
In this appendix, we will derive Eq. \eqref{surv-eta-1-eq-7} which describes $\langle T_{r}(x_0,v_0) \rangle$ as $r \to \infty$. We rewrite the expression of $\alpha (r)$ from Eq. \eqref{surv-eta-1-eq-2},
\begin{align}
\alpha (r)=\int_0^{\infty}\frac{dy}{y^{5/3}} e^{-y x_0} \text{Ai}\left( v_0 y^{1/3}+\frac{r}{y^{2/3}} \right) \left[ 1+\frac{1}{4 \sqrt{\pi}} \Gamma \left( -\frac{1}{2},\frac{2 r^{3/2}}{3 y} \right)\right].
\label{MFPT-large-gam-eq-1}
\end{align}
Changing variable $y x_0=z$, we get
\begin{align}
\alpha (r)=x_0^{2/3}\int_0^{\infty}\frac{dz}{z^{5/3}} e^{-z} \text{Ai}\left( \frac{v_0}{x_0^{1/3}} z^{1/3}+\frac{r x_0^{2/3}}{z^{2/3}} \right) \left[ 1+\frac{1}{4 \sqrt{\pi}} \Gamma \left( -\frac{1}{2},\frac{2 x_0r^{3/2}}{3 z} \right)\right].
\label{MFPT-large-gam-eq-2}
\end{align}
Looking at the expression above, we realise that for $r >>x_0^{-2/3}$ and $r>>v_0^2$, the arguments of Airy function and incomplete gamma function are very large. In this limit we can use the asymptotic forms of Airy function and incomplete gamma function. The asymptotic forms read as,
\begin{align}
&\text{Ai}(\omega) \simeq \frac{1}{2 \sqrt{\pi} \omega ^{1/4}} e^{-\frac{2}{3}\omega^{3/2}},~~\omega \to \infty \nonumber \\
&\Gamma \left( -\frac{1}{2}, \omega\right) \simeq \frac{e^{-\omega}}{\omega ^{3/2}},~~~~~~~~~\omega \to \infty.
\label{MFPT-large-gam-eq-3}
\end{align}
Using these forms in Eq. \eqref{MFPT-large-gam-eq-2} and performing the integration over $z$, we get in the leading order for $r >>x_0^{-2/3}$ and $r>>v_0^2$,
\begin{align}
\alpha (r) \simeq \frac{1}{r} \sqrt{\frac{3}{8}}~ \text{exp} \left( -v_0 \sqrt{r} - \sqrt{\frac{8}{3} x_0 r^{3/2}}\right).
\label{MFPT-large-gam-eq-4}
\end{align}
Inserting this form of $\alpha (r)$ in Eq. \eqref{surv-eta-1-eq-6}, one gets $\langle T_{r}(x_0,v_0)\rangle$ in Eq. \eqref{surv-eta-1-eq-7}.

\end{document}